\documentclass[a4paper,11pt]{article}
\usepackage{jheppub} 

\usepackage{graphicx}%
\graphicspath{{Figure/}}

\usepackage{adjustbox}
\usepackage{amsmath}

\usepackage{xspace}
\usepackage{siunitx}

\usepackage{placeins}

\makeatletter
\renewcommand{\section}{\@startsection{section}{1}{\z@}{10pt}{1pt}{\large\bfseries}}
\renewcommand{\subsection}{\@startsection{subsection}{2}{\z@}{5pt}{1pt}{\normalsize\bfseries}}
\makeatother

\usepackage[italic,italicGreek]{heppennames}

\DeclareSIUnit{\mrad}{mrad}

\newcommand{\PAlepton}{\ensuremath{\overline{l}}}

\newcommand{\hc}[1]{{}^\ddagger #1}
\newcommand*{\tmp}[4]{\ensuremath{
    {#4
    \ifx\empty#3\empty\ifx\empty#1\empty\else^{#1}\fi\else^{#1(#3)}\fi
    \ifx\empty#2\empty\else_{#2}\fi}
}}
\newcommand*{\qq }[4][]{\tmp{#2}{#3}{#4}{#1{O}}}

\newcommand{\FDF}{(\varphi^\dagger i\!\!\overleftrightarrow{D}_\mu\varphi)}
\newcommand{\FDFI}{(\varphi^\dagger i\!\!\overleftrightarrow{D}^I_\mu\varphi)}

\newcommand{\sW}{\ensuremath{s_{\mathrm{W}}}\xspace}
\newcommand{\cW}{\ensuremath{c_{\mathrm{W}}}\xspace}

\newcommand*{\eftOp}[4]{\ensuremath{
    {#4
    \ifx\empty#3\empty\ifx\empty#1\empty\else^{#1}\fi\else^{#1(#3)}\fi
    \ifx\empty#2\empty\else_{#2}\fi}
}}

\newcommand{\PQqM} {\eftOp{-}{\varphi Q}{}{c}\xspace}
\newcommand{\PQqa} {\eftOp{3}{\varphi Q}{}{c}\xspace}
\newcommand{\cpt}  {\eftOp{}{\varphi \PQt}{}{c}\xspace}

\newcommand{\ctW}  {\eftOp{}{ \PQt\PW}{}{c}\xspace}
\newcommand{\ctWI} {\eftOp{I}{ \PQt\PW}{}{c}\xspace}
\newcommand{\ctZ}  {\eftOp{}{ \PQt\PZ}{}{c}\xspace}
\newcommand{\ctZI} {\eftOp{I}{ \PQt\PZ}{}{c}\xspace}

\newcommand{\cQla} {\eftOp{3}{Ql}{1}{c}\xspace}
\newcommand{\cQlM} {\eftOp{-}{Ql}{1}{c}\xspace}
\newcommand{\cQe}  {\eftOp{}{Q\Pe}{1}{c}\xspace}
\newcommand{\ctl}  {\eftOp{}{ \PQt l}{1}{c}\xspace}
\newcommand{\cte}  {\eftOp{}{ \PQt \Pe}{1}{c}\xspace}
\newcommand{\ctlS} {\eftOp{S}{ \PQt}{1}{c}\xspace}
\newcommand{\ctlSI}{\eftOp{SI}{ \PQt}{1}{c}\xspace}
\newcommand{\ctlT} {\eftOp{T}{ \PQt}{1}{c}\xspace}
\newcommand{\ctlTI}{\eftOp{TI}{ \PQt}{1}{c}\xspace}

\newcommand{\cQlal} {\eftOp{3}{Ql}{\ell}{c}\xspace}
\newcommand{\cQlMl} {\eftOp{-}{Ql}{\ell}{c}\xspace}
\newcommand{\cQel}  {\eftOp{}{Q\Pe}{\ell}{c}\xspace}
\newcommand{\ctll}  {\eftOp{}{ \PQt l}{\ell}{c}\xspace}
\newcommand{\ctel}  {\eftOp{}{ \PQt \Pe}{\ell}{c}\xspace}
\newcommand{\ctlSl} {\eftOp{S}{ \PQt}{\ell}{c}\xspace}
\newcommand{\ctlSIl}{\eftOp{SI}{ \PQt}{\ell}{c}\xspace}
\newcommand{\ctlTl} {\eftOp{T}{ \PQt}{\ell}{c}\xspace}
\newcommand{\ctlTIl}{\eftOp{TI}{ \PQt}{\ell}{c}\xspace}

\newcommand*{\GeV}{\xspace\ensuremath{\text{GeV}}\xspace}

\newcommand*{\ttbar}{\ensuremath{t\bar{t}}\xspace}

\newcommand*{\WHIZ}{{\tt WHIZARD3}\xspace}

\newcommand*{\PYTHIAs}{{\tt PYTHIA6}\xspace}
\newcommand*{\PYTHIAe}{{\tt PYTHIA8}\xspace}

\newcommand*{\DELPHES}{{\tt DELPHES}\xspace}

\newcommand*{\MADGRAPH}{{\tt MadGraph\_aMC@NLO}\xspace}

\newcommand{\EFT}{\text{SMEFT}}
\newcommand{\SM}{\text{SM}}
\newcommand{\reco}{\text{reco}}

\newcommand*{\sqrts}{\ensuremath{\sqrt{s}}\xspace}

\newcommand*{\sqrtsTop}{\ensuremath{\sqrts = \SI{365}{\GeV}}\xspace}

\newcommand{\WbWb}{\ensuremath{W^+bW^-\bar{b}}}
\newcommand{\bbbar}{\ensuremath{b\bar{b}}}
\newcommand{\qqbar}{\ensuremath{q\bar{q}}}

\preprint{IRMP-CP3-26-21}

\title{%
Top-quark pair production sensitivity to heavy new physics at the FCC-ee
}

\author[a,b]{Jacopo De Piccoli,}
\author[c]{Gauthier Durieux,}
\author[d]{Sergio S\'anchez Cruz,}
\author[e]{and Michele Selvaggi}

\affiliation[a]{Istituto Nazionale di Fisica Nucleare - Sezione di Padova, via Marzolo 8, Padova, 35131, Italy}
\affiliation[b]{Dipartimento di Fisica e Astronomia Galileo Galilei, Universit\`a di Padova, Padova, Italy}
\affiliation[c]{Center for Cosmology, Particle Physics and Phenomenology (CP3), University of Louvain (UCLouvain),
Chemin du Cyclotron 2, B-1348 Louvain-la-Neuve, Belgium
}
\affiliation[d]{Departamento de F\'isica and ICTEA, Universidad de Oviedo, Oviedo, 33007, Spain}
\affiliation[e]{European Organisation for Nuclear Research (CERN), Geneva, Switzerland}

\emailAdd{jacopo.de.piccoli@cern.ch}
\emailAdd{sergio.sanchez.cruz@cern.ch}
\emailAdd{gauthier.durieux@uclouvain.be}
\emailAdd{michele.selvaggi@cern.ch}

\abstract{%
We study the sensitivity of top-quark pair production at the FCC-ee to heavy new physics, parametrised in the Standard Model effective field theory, using discriminants based on matrix elements. The analysis includes detector reconstruction effects, obtained with a fast simulation of the IDEA detector concept, and jet flavour tagging. We consider dimension-six operators that modify the electroweak couplings of the top quark or generate four-fermion interactions, at leading order and in diagrams with two resonant top quarks.
The full six-body kinematic information is employed to maximise the discrimination power against the Standard Model.
Two bins per dimension in the space of 12 discriminants, one per operator considered, provide a simple and robust methodology to set nearly optimal multidimensional constraints.
Bounds on the new physics scale can reach values up to 30 TeV in a single-operator fit and up to 5 TeV in a global multi-operator fit.
A combination with High-Luminosity LHC prospects on top-quark operators shows complementary sensitivities which help lift approximate degeneracies.
}

\usepackage{newunicodechar}
\newunicodechar{−}{-}

\begin{document}
\maketitle
\flushbottom



\section{Introduction} \label{sec:intro}

As the heaviest known elementary particle, the top quark is the only fermion with a mass close to the electroweak symmetry breaking scale, making it a prime candidate for probing physics beyond the Standard Model (SM).
The SM effective field theory (SMEFT) systematically parametrises the lower-energy effects of heavy new physics through higher-dimensional operators added to the SM Lagrangian.
A subset of these operators induce modifications of top-quark interactions.
They first arise at mass-dimension six, while higher-dimensional ones are expected to be further suppressed.
Isolating their effects in top-quark observables would provide indirect evidence for new physics at higher energy scales.

The couplings of the top quark have been extensively characterised at hadron colliders, where top-quark production dominantly proceeds in pairs, through strong interactions.
Therefore, it mostly probes the $t\bar{t}g$ and $t\bar{t}q\bar{q}$ couplings generated by dimension-six SMEFT operators.
On the contrary, top-quark decay and single production are weak processes which give access to charged-current $t\bar{b}W$, $t\bar{b}q\bar{q}'$, or $t\bar{b}\ell\bar{\nu}$ interactions.
Neutral-current interactions like $t\bar{t}\gamma$, $t\bar{t}Z$, $t\bar{t}\ell\bar{\ell}$ and $t\bar{t}H$ are probed through much rarer associated production processes.
In particular, a number of measurements of these processes have been performed at the LHC~\cite{CMS:2023xyc,CMS:2025dpp,ATLAS:2025yww,ATLAS:2025adk,ATLAS:2023eld,CMS:2019too,CMS:2022lmh,ATLAS:2024hmk,Celada:2024mcf,deBlas:2025xhe}.
However, High-Luminosity LHC (HL-LHC) projections indicate that several directions in the coupling parameter space will remain unconstrained~\cite{Collaboration:2938605}, strongly motivating complementary studies at future lepton colliders.

Future lepton colliders such as the FCC-ee offer a more direct way to probe the top-quark electroweak couplings, since top-quark pair production proceeds exclusively through electroweak interactions.
In addition, the clean experimental conditions of $e^+e^-$ collisions at $\sqrt{s} = 340-365\text{ GeV}$ allow for a full reconstruction of the top-antitop ($t\bar{t}$) system.
This precision enables direct access to spin correlations and angular distributions that are difficult to exploit at hadron machines.
This clean environment also translates into significantly lower systematic uncertainties, particularly those related to jet-flavour tagging and luminosity.

The present paper focuses on the FCC-ee sensitivity to SMEFT in $e^+e^-\to t\,\bar{t}$ production using a fast simulation of the IDEA detector concept and modern transformer-based jet-flavour tagging.
Background contamination and reconstruction effects are thereby incorporated directly in the construction of the observables, going beyond the parton-level treatment of earlier studies~\cite{Janot:2015yza,Durieux:2018tev}.

We exploit discriminants based on matrix elements to maximise the sensitivity to SMEFT operators, exploiting the full six-body kinematics of the reconstructed $\PQb\PAQb\ell\nu\PQq\PQq'$ final state.
A minimal multidimensional binning of the discriminant distributions is found to be practical to implement while providing robust results and nearly optimal sensitivity.

Earlier analyses of beyond-the-SM (BSM) deformations of $e^+e^-\to t\,\bar{t}$ production have exploited various fractions of the available kinematic information through statistically optimal observables~\cite{Atwood:1991ka, Grzadkowski:2000nx, Janot:2015yza, Khiem:2015ofa, Durieux:2018tev}.
In the limit where differential rate only depends linearly on the parameters of interest, a discrete set of such matrix-element-based observables (one per parameter of interest) can be shown to minimise the volume of statistical constraints.
Using the statistically optimal observables defined in Ref.~\cite{Grzadkowski:2000nx} from the energy and production angle of one final-state charged lepton, Ref.~\cite{Janot:2015yza} notably studied the FCC-ee sensitivity to anomalous $t\bar{t}Z$ and $t\bar{t}\gamma$ form factors.
The experimental aspects of such measurements, in particular the background rejection and the migrations effects on the reconstruction of the top-quark direction were studied in detail with a full simulation of the ILD detector concept at the ILC~\cite{Amjad:2013tlv,Amjad:2015mma}.
The present work propagates reconstruction effects directly into discriminants built from the full six-body kinematics, and quantifies their impact on the simultaneous determination of all operator coefficients, for which no detector level study exists at the FCC-ee.
Optimal observables built from the $\WbWb$ kinematic information were subsequently employed in Ref.~\cite{Durieux:2018tev} to study the sensitivity of future lepton collider projects to SMEFT operator coefficients.
In addition to CP-even and -odd modifications of $t\bar{t}Z$ and $t\bar{t}\gamma$ interactions, four-fermion operators of $e^+e^-t\,\bar{t}$ type were also included.
These results currently serve as input to several global and top-specific SMEFT analyses of the reach of future lepton colliders (see e.g.\ Refs.~\cite{Cornet-Gomez:2025jot, Armadillo:2026mvp}).

The rest of this paper is organised as follows: Section~\ref{sec:smeft} defines the top-quark SMEFT operators considered; Section~\ref{sec:eventsim} describes the event generation, simulation, and experimental reconstruction as well as the optimal discriminant methodology; Section~\ref{sec:results} summarises the results of the single and multi-dimensional fits; conclusions and outlook are provided in Section~\ref{sec:Conclusion}.


\section{Operators considered}
\label{sec:smeft}

We consider the operators involving a top quark that contribute at tree level to top-quark pair production ($e^+e^- \to t\,\bar{t}$) at the FCC-ee and adopt the LHC TOP WG notation and conventions~\cite{AguilarSaavedra:2018nen} for those.
Two main groups of operators can be distinguished:
\begin{itemize}

\item Operators involving two quarks and bosons modify the couplings of the top and bottom quarks to the electroweak gauge bosons.
Among these, the $t\bar{t}Z$ and $t\bar{t}\gamma$ couplings enter $e^+e^- \to t\,\bar{t}$ production at tree level.
Dipolar interactions have both CP-even and -odd components, respectively proportional to the real and imaginary parts of operator coefficients, while modifications of the $t\bar{t}Z$ vector current can affect top quarks of either left- or right-handed chiralities.

\item Operators involving two quarks and two leptons generate $e^+e^-t\,\bar{t}$ four-fermion contact interactions which give rise to $e^+e^- \to t\,\bar{t}$ amplitudes growing quadratically with the centre-of-mass energy $\sqrt{s}$.

\end{itemize}

Following Ref.~\cite{AguilarSaavedra:2018nen}, we assume flavour diagonality in the lepton sector ($[U(1)_{l+e}]^3$) and a baseline $U(2)_q \times U(2)_u \times U(2)_d$ flavour symmetry among the light-quark generations.
The resulting degrees of freedom are listed in Table~\ref{tab:operators}.
In this parametrisation, $\PQqa$ and $\cQla$ are not constrained by $e^+e^- \to t\,\bar{t}$ production.
The most powerful probe for these degrees of freedom would be $e^+e^-\to b\,\bar{b}$ production on and above the $Z$ pole.
Since this process is not included in our analysis, we simply omit $\PQqa$ and $\cQla$.

We also note that the scalar and tensor four-fermion operators, with $\ctlS$ and $\ctlT$ coefficients, produce $e^+e^- \to t\,\bar{t}$ amplitudes where the electron-positron pair arises in helicity combinations different from the SM ones.
Their interference with SM amplitudes is therefore suppressed by the electron mass, which we neglect.
As a result, the dependence of the differential rate on $\ctlS$ and $\ctlT$ is purely quadratic, stemming exclusively from the modulus squared of diagrams featuring an insertion of these operators.
Although the interference between scalar and tensor amplitudes could differ for real and imaginary coefficients, we find these effects to be numerically negligible and therefore do not show results for the imaginary parts $\ctlSI$ and $\ctlTI$.

The degrees of freedom explicitly considered in our analysis are therefore the $12$ following ones:
\begin{equation}\begin{gathered}
\PQqM,\ \cpt,\\
\ctW,\ \ctZ,\\
\ctWI,\  \ctZI,
\end{gathered}
\qquad
\begin{gathered}
\cQlM,\ \ctl,\\
\cQe,\ \cte,\\
\ctlS,\ \ctlT\,.
\end{gathered}\end{equation}

\begin{table}[tb]
\renewcommand{\arraystretch}{1.3}
\centering
\begin{tabular}[tb]{p{2cm} p{6cm} p{5.8cm}}
\hline
\textbf{Operator} & \textbf{Definition} & \textbf{Coefficient} \\
\hline
$\qq{1}{\varphi \PQq}{ij}$ & $\FDF (\PAQq_i\gamma^\mu \PQq_j)$ & $\PQqM + \PQqa$ \\ 
$\qq{3}{\varphi \PQq}{ij}$ & $\FDFI (\PAQq_i\gamma^\mu\tau^I \PQq_j)$ & $\PQqa$ \\ 
$\qq{}{\varphi  \PQu}{ij}$ & $\FDF (\PAQu_i\gamma^\mu  \PQu_j)$ & $\cpt$ \\ 
$\hc{\qq{}{ \PQu\PW}{ij}}$ & $(\PAQq_i\sigma^{\mu\nu}\tau^I \PQu_j)\:\tilde{\varphi}\: \PW_{\mu\nu}^I$ & $\ctW + i \ctWI$ \\ 
$\hc{\qq{}{ \PQu\PB}{ij}}$ & $(\PAQq_i\sigma^{\mu\nu} \PQu_j)\:\tilde{\varphi}\: \PB_{\mu\nu}$ & $\frac{\cW \ctW - \ctZ}{\sW} + i \frac{\cW \ctWI - \ctZI}{\sW}$    \vspace{3 pt} \\
\hline
$\qq{1}{l\PQq}{ijkm}$ & $(\PAlepton_i\gamma^\mu l_j)(\PAQq_k\gamma^\mu \PQq_m)$ & $\cQlMl + \cQlal$ \\ 
$\qq{3}{l\PQq}{ijkm}$ & $(\PAlepton_i\gamma^\mu \tau^I l_j)(\PAQq_k\gamma^\mu \tau^I \PQq_m)$ & $\cQlal$ \\ 
$\qq{}{l \PQu}{ijkm}$ & $(\PAlepton_i\gamma^\mu l_j)(\PAQu_k\gamma^\mu  \PQu_m)$ & $\ctll$ \\ 
$\qq{}{\Pe\PAQq}{ijkm}$ & $(\overline{\Pe}_i\gamma^\mu \Pe_j)(\PAQq_k\gamma^\mu \PQq_m)$ & $\cQel$ \\ 
$\qq{}{\Pe \PQu}{ijkm}$ & $(\overline{\Pe}_i\gamma^\mu \Pe_j)(\PAQu_k\gamma^\mu  \PQu_m)$ & $\ctel$ \\ 
$\hc{\qq{1}{l\Pe\PQq \PQu}{ijkm}}$ & $(\PAlepton_i \Pe_j)\;\varepsilon\;(\PAQq_k  \PQu_m)$ & $\ctlSl + i \ctlSIl$ \\ 
$\hc{\qq{3}{l\Pe\PQq \PQu}{ijkm}}$ & $(\PAlepton_i \sigma^{\mu\nu} \Pe_j)\;\varepsilon\;(\PAQq_k \sigma_{\mu\nu}  \PQu_m)$ & $\ctlTl + i \ctlTIl$ \\
\hline
\end{tabular}
\caption{Two- and four-fermion operators considered.
Quark and lepton generation indices are set to the third and first generation, respectively.
The sine and cosine of the weak mixing angle are abbreviated as $s_W$ and $c_W$, respectively.
}
\label{tab:operators}
\end{table}


\section{Experimental analysis strategy } \label{sec:eventsim}

In this section, we outline an experimental strategy to probe the coefficients of the operators introduced in the previous section in FCC-ee runs at, and above, the top pair production threshold.
We consider a simplified run scenario compared to the ten-point threshold scan proposed in Ref.~\cite{Defranchis:2025auz}, using a dataset of $2.72~\mathrm{ab}^{-1}$ recorded at $\sqrt{s} = 365~\mathrm{GeV}$, with the possibility of collecting additional $0.2~\mathrm{ab}^{-1}$ at both $\sqrt{s} = 340~\mathrm{GeV}$ and $345~\mathrm{GeV}$ to gain a handle on the $\sqrt{s}$ dependence of SMEFT effects.
We do not expect our final results to depend significantly on the exact repartition of the integrated luminosity across threshold centre-of-mass energies.
The correlations between measurements performed by the four different experiments around the FCC-ee ring are neglected and the luminosities quoted above are simply the sum over those collected at the four interactions points.

The sensitivity to the operators is estimated by the means of samples of simulated events, which are also described in this section.

\subsection{Event generation and matrix-element reweighting}

The signal process is defined as $e^{+}e^{-}\to \WbWb$ with one $W$ boson decaying into 
an electron or muon and the corresponding neutrino, and the other $W$ boson decaying hadronically, thereby producing a $\PQb\PAQb\ell\nu\PQq\PQq'$ final state.
The remaining $\WbWb$ decay modes, namely the fully leptonic, fully hadronic, and the channels involving $\tau$ leptons, are treated as a separate background category (other $\WbWb$).
The $\WbWb$ matrix element computed in the SM also includes single-top and non-resonant contributions; all are part of the signal definition in this analysis.
The dominant additional backgrounds are inclusive $WW$, $ZZ$, $WWZ$ ($Z \to \bbbar$), and $\qqbar$ production.
Small interference effects between the hadronic final states of $WW$ and $ZZ$ production and photon-induced multi-jet production are neglected, as is the interference between $\WbWb$ and $WWZ$ ($Z \to \bbbar$), consistent with the approach followed in Ref.~\cite{Defranchis:2025auz}.
The $WWH$ ($H \to \bbbar$) background is not considered due to its significantly smaller production cross section.

The $\WbWb$, $WWZ$, and $\qqbar$ samples are produced with the \WHIZ generator~\cite{Kilian:2007gr} interfaced with \PYTHIAs~\cite{Sjostrand:2006za}, while $ZZ$ and $WW$ are generated with {\tt PYTHIA8}~\cite{Sjostrand:2014zea}.
All samples are produced at leading order and are normalised to the cross sections used in Ref.~\cite{Defranchis:2025auz} which are summarised in Table~\ref{tab:MC_samples_eft}.

\begin{table}[tb]
    \centering
    \small
    \begin{tabular}{lllccc}
    	\hline
        \multicolumn{1}{l}{Process} & \multicolumn{1}{l}{Decays} & \multicolumn{1}{l}{Generator} & \multicolumn{3}{c}{Cross section [pb]} \\
        \multicolumn{3}{c}{} & $340~\GeV$ & $345~\GeV$ & $365~\GeV$ \\
        \hline
        $\WbWb$ (signal)        & $e/\mu\,\nu\,jj$    & \WHIZ + \PYTHIAs & 0.029 & 0.144 & 0.144 \\
        $\WbWb$ (others)        & remaining           & \WHIZ + \PYTHIAs & 0.071 & 0.356 & 0.356 \\
        $WWZ$                   & $Z\to \bbbar$       & \WHIZ + \PYTHIAs & $1.32\times 10^{-3}$ & $1.46\times 10^{-3}$ & $2.02\times 10^{-3}$ \\
        $\qqbar$                & n.a.                & \WHIZ + \PYTHIAs & 26.3 & 25.6 & 22.8 \\
        $ZZ$                    & inclusive           & \PYTHIAe & 0.932 & 0.916 & 0.643 \\
        $WW$                    & inclusive           & \PYTHIAe & 12.1 & 11.9 & 10.7 \\
        \hline
    \end{tabular}
    \caption{Monte Carlo samples and total cross sections for the signal and background processes, at the three centre-of-mass energies of interest.}
    \label{tab:MC_samples_eft}
\end{table}

The contributions of the various SMEFT operators to the final state are accounted for using event reweighting~\cite{Mattelaer:2016gcx}.
The weights are determined as the ratio of matrix elements under different hypotheses and are applied to the nominal \WHIZ samples.
The matrix elements are computed with the \MADGRAPH generator~\cite{Alwall:2011uj} for 
all the $\WbWb$ decay channels, including only diagrams with two resonant top quarks.
They are functions of the four-momenta of the incoming electron and positron, the \PQb quarks, and the leptons and quarks produced in the decay of the \PW bosons.
The computation is performed parametrically as a function of the operator coefficients.
The contribution from the different helicity configurations are summed when computing the matrix element to increase the statistical power of the reweighting procedure, as discussed in Ref.~\cite{Belvedere:2024nzh}.

\begin{figure}
\centering
\includegraphics[width=0.49\textwidth]{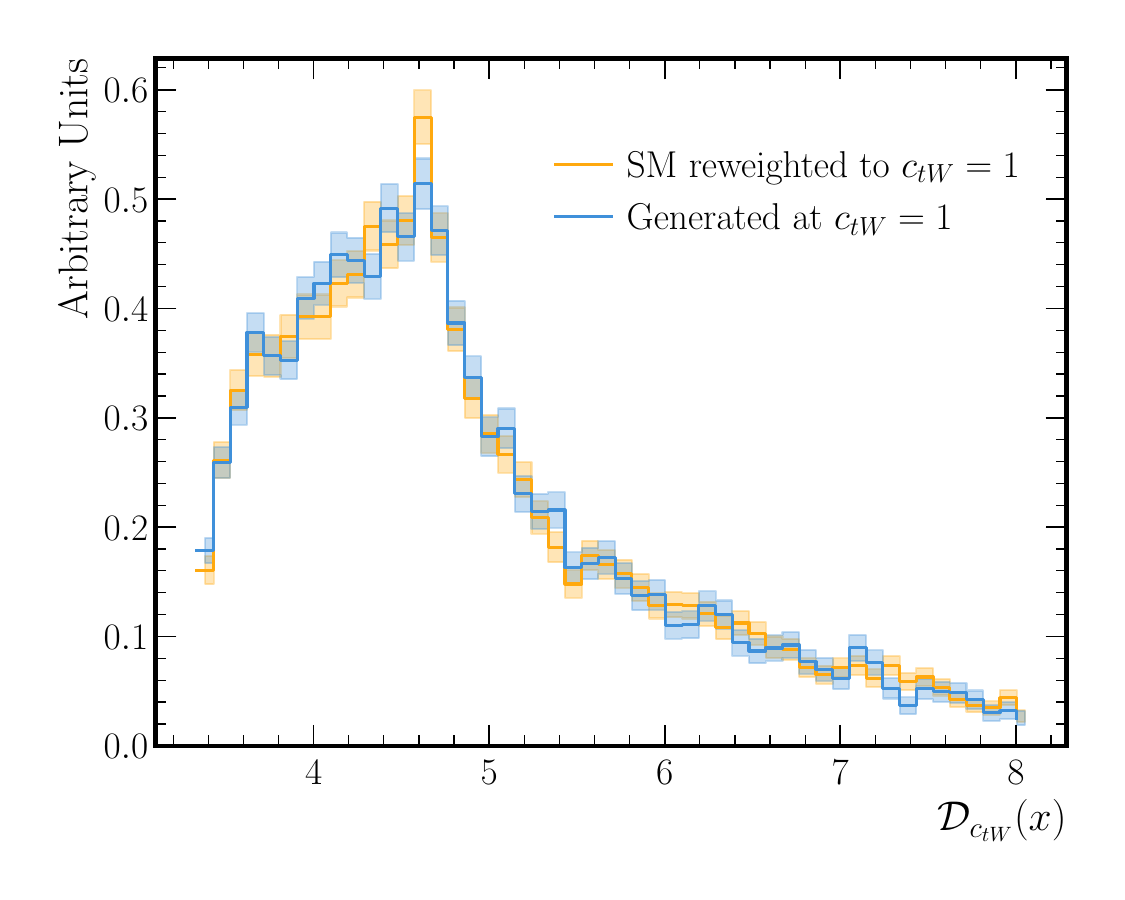}
\includegraphics[width=0.49\textwidth]{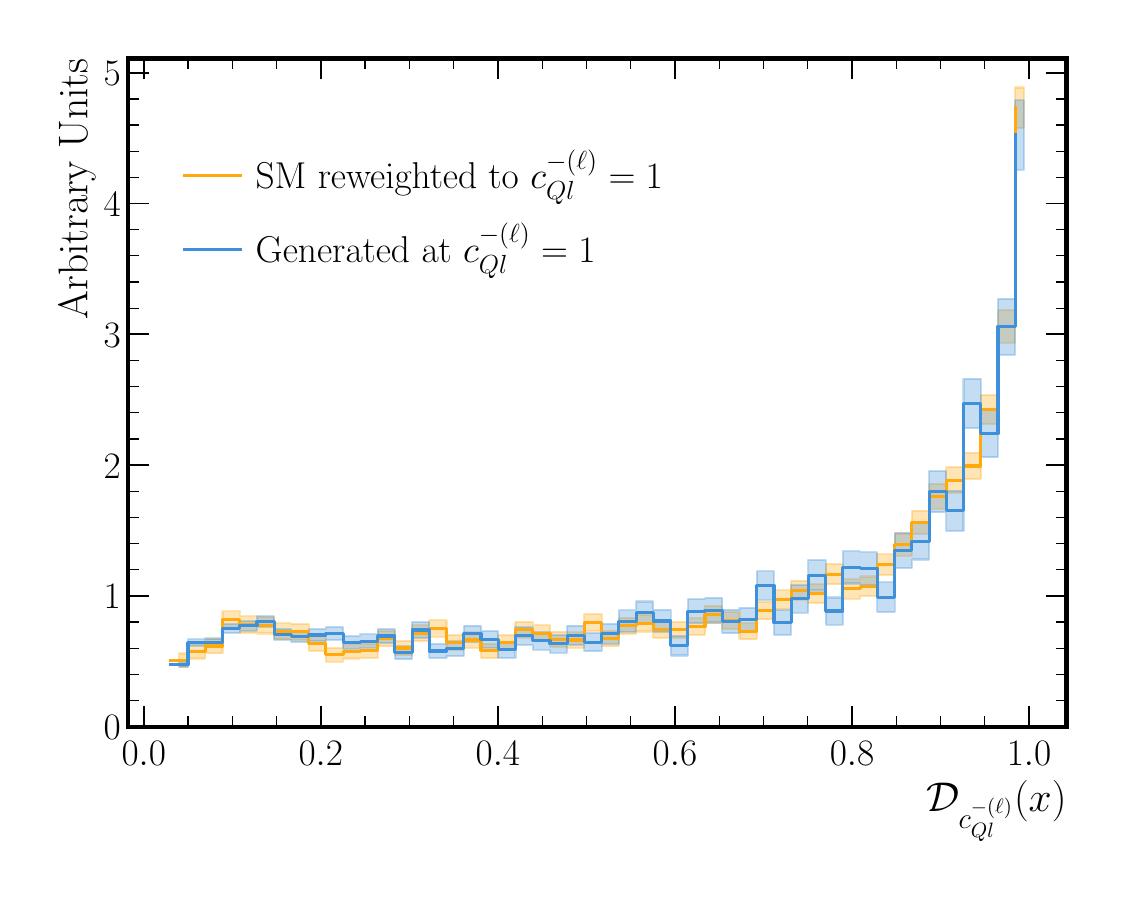}
\caption{Distributions of the $\mathcal{D}_{\ctW}$ (left) and  $\mathcal{D}_{\cQlM}$ (right) discriminants computed with generator-level \MADGRAPH samples.
The orange lines are obtained from samples generated directly with non-zero values for the operator coefficient ($c=1$ with $\Lambda=1\:$Tev), while the blue lines are obtained from reweighted SM samples.
The statistical uncertainties are shown as shaded bands.
}
\label{fig:rw_validation}
\end{figure}

This procedure has been validated using dedicated samples generated for different values of the operator coefficients, with the \MADGRAPH generator.
The distribution of the optimal observables $\mathcal{D}$ described in Section~\ref{sec:oo} was compared between samples generated assuming non-zero coefficients and samples generated under the SM hypothesis and appropriately reweighted.
Figure~\ref{fig:rw_validation} shows the good agreement obtained e.g.\ for the \ctW and \cQlM operators.
This validation process has been successfully carried out for all the considered operators.

\subsection{Event reconstruction and selection}

Detector effects are simulated using the \DELPHES~\cite{deFavereau:2013fsa} 
parametrisation of the IDEA detector concept~\cite{IDEAStudyGroup:2025gbt} at the FCC-ee.
Final-state particles are reconstructed using the particle-flow algorithm implemented in \DELPHES, which combines tracking and calorimetric information to provide an optimal estimate of the momentum of each visible particle.
The missing momentum is computed as the recoil against the sum of all visible reconstructed particles, both charged and neutral.

Events are required to contain exactly one isolated lepton (electron or muon) with transverse momentum $p_T > 15~\GeV$ and cosine of its polar angle $|\cos\theta| < 0.97$.
The isolation criterion $I < 0.2$ is computed as the scalar sum of the momenta of charged particles within an angular distance $0.01 < \Delta R < 0.2$ around the lepton direction, relative to the lepton momentum.
This requirement selects the semi-leptonic final state targeted by this analysis and provides strong discrimination against the fully hadronic $\WbWb$ and $\qqbar$ (including $\bbbar$) backgrounds.

The selected isolated lepton is then removed from the list of reconstructed particles, and the remaining particles are clustered into jets using the generalised $k_T$ algorithm~\cite{Catani:1991hj} implemented in \textsc{FastJet}~\cite{Cacciari:2011ma}, used in anti-$k_T$ mode with a radius parameter $R = 0.5$.
Jet flavour tagging probabilities are assigned with a transformer-based tagger~\cite{Bedeschi:2022rnj, aumiller_2024_8g834-jv464}.

At least four jets are required, with the four leading jets satisfying 
$|\cos\theta| < 0.97$ and $p_T > 15~\GeV$.
The missing momentum satisfies similar requirements, i.e.\ $|\cos\theta| < 0.97$ and $p_T^{\rm{miss}} > 10~\GeV$, consistent with the presence of a neutrino from the leptonic $W$ decay and providing additional rejection against $\qqbar$ and $\bbbar$ events, where the missing momentum originates only from semi-leptonic heavy-flavour decays and tends to be small.
A minimum charged hadron multiplicity $N_{\text{ch}} > 20$ is also imposed, in order to suppress non-hadronic final states such as $\tau$-pair events or radiative $\gamma^*/Z^* \to e^+e^-/\mu^+\mu^-$ events.

\begin{figure}[tb]
\centering
\includegraphics[width=0.48\linewidth]{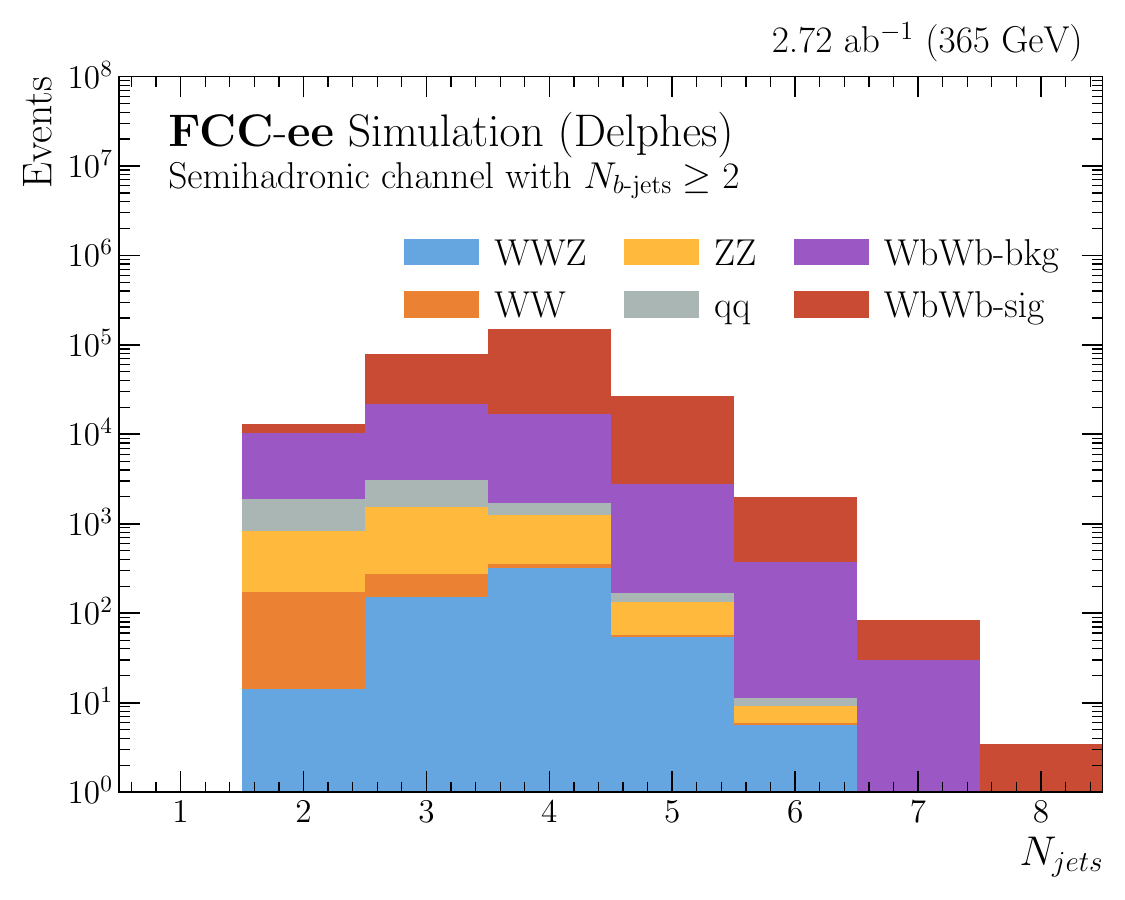}    \includegraphics[width=0.48\linewidth]{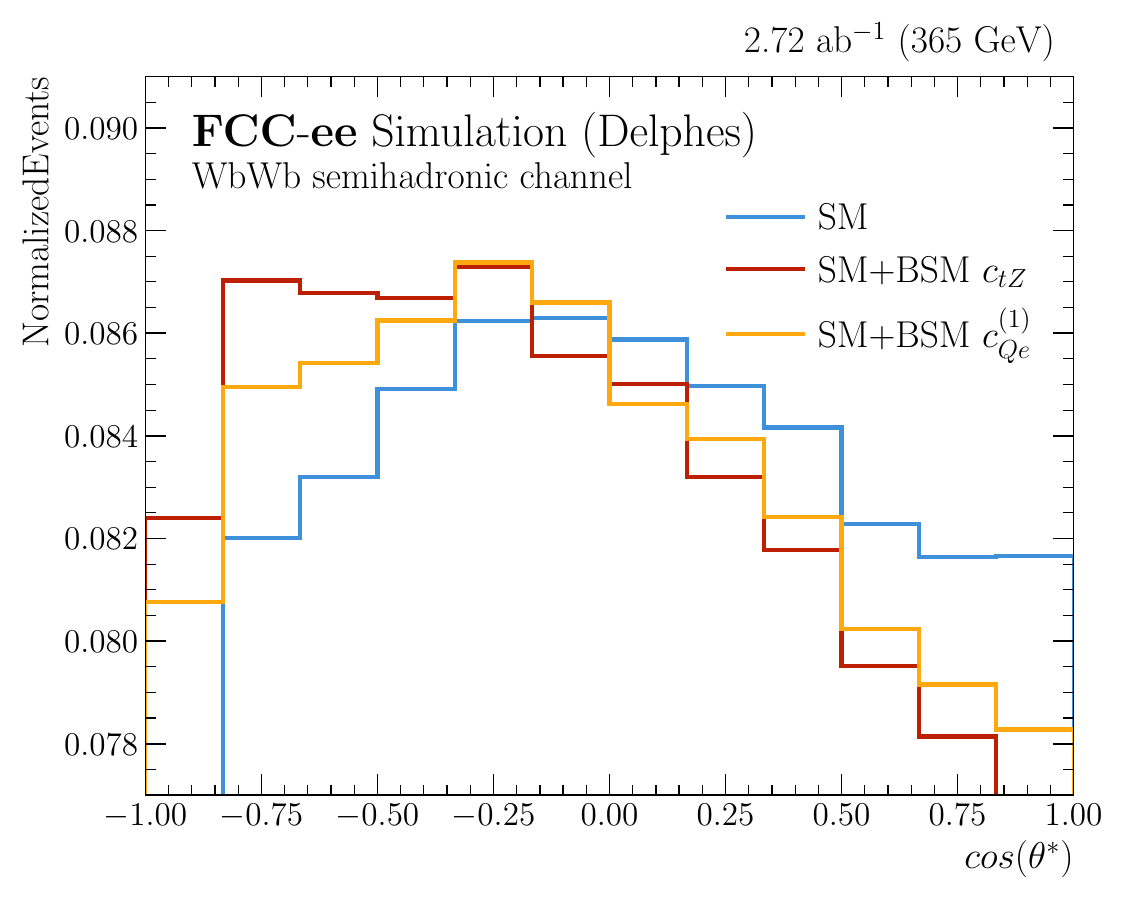}
\hfill
\caption{Left: expected yield distributions for signal and background processes after selecting events with one isolated lepton and two $b$-jets.
Right: normalised distribution of the top polarisation angle $\cos\theta^*$ for the SM dataset and its reweightings to $\ctZ/\Lambda^2=1\:\text{TeV}^{-2}$ or $\cQe/\Lambda^2=1\:\text{TeV}^{-2}$.}
\label{fig:selection}
\end{figure}

The flavour tagger output is then used to identify jets originating from the
hadronisation of $b$ quarks.
Jets are considered $b$-tagged if their score
exceeds 0.9, corresponding to a $b$-tagging efficiency of 79\%.
Events are selected by requiring at least two $b$-tagged jets among the four leading ones.
The jet multiplicity in selected events (with the minimum jet multiplicity requirement momentarily relaxed to two) is shown in Figure~\ref{fig:selection} (left). 
The signal populates predominantly the $N_{jets} = 4$ category, as expected from the presence of two $b$ quarks and two light quarks in the final state. In contrast, the $WW$, $ZZ$, and $\qqbar$ backgrounds feature lower jet multiplicities.
Requiring a minimum of four jets of which at least two are $b$-tagged, provides clean separation against these processes and a high signal efficiency.

After the full event selection, the efficiency at $\sqrt{s} = 365~\GeV$ is 41.1\% for the $\WbWb$ signal in the $e/\mu$ semi-leptonic channel, 1.9\% for the other $\WbWb$ final states, 6.8\% for $WWZ$ and $<0.1\%$ for the remaining backgrounds.
Comparable values are obtained at $\sqrt{s} = 340$ and $345~\GeV$.
The final event sample at $\sqrt{s} = 365~\GeV$ contains 180861 events, of which 160830 ($88.9\%$) are signal semi-leptonic $\WbWb$ events in the $e/\mu$ channel, 18271 ($10.1\%$) are other $\WbWb$ final states, and 1853 ($1.0\%$) come from the remaining backgrounds.
The contamination from processes other than $WbW\bar{b}$ is therefore  $\mathcal{O}(1\%)$ and is expected to be controlled at the sub-percent level.
These background contributions are therefore neglected in the following.
This level of background contamination is comparable to that of the analysis presented in Ref.~\cite{Defranchis:2025auz} using a similar framework, which adopted a parametric $b$-tagging approach.

Figure~\ref{fig:selection} (right) shows the distribution of $\cos\theta^*$, the angle between the final-state lepton and the top quark, in the top rest frame. 
This observable is sensitive to the top-quark polarisation and discriminates between the SM and nonzero operator coefficients.
However,
different operators tend to affect kinematic distributions in a similar way, which limits the ability of any single distribution to disentangle them.
This motivates the use of discriminants based on matrix elements, which exploit the full kinematic information of the event.

\subsection{Matrix-element-based discriminants}
\label{sec:oo}

In order to maximise the sensitivity to SMEFT contributions, we employ matrix-element-based discriminants as a proxy for the per-event likelihood ratio, parametrised by the operator coefficients.
For each event, we compute a discriminant as follows
\begin{equation}
\mathcal{D}(x_{\reco},\{c_i\}) =\frac{|A_{\SM}(x_{\reco})+\sum_iA_{\EFT,i}(x_{\reco}) \: c_i/\Lambda^2  |^2 }{|A_{\SM}(x_{\reco})|^2}
\label{eq:discriminant}
\end{equation}
where $A_{\SM}$ and $A_{\EFT,i}$ are the SM and SMEFT amplitudes for $e^+e^-\to\PQt\PAQt\to\PQb\PAQb\ell\nu\PQq\PQq'$ production.
The values of $c_i/\Lambda^2$ operator coefficients determine at which point of the SMEFT parameter space the discriminant is optimal and $x_{\reco}$ collectively denotes the reconstructed kinematics of the final state particles, i.e.\ their four-momenta.

For reconstructed events the incoming electron and positron are assigned the nominal beam
four-momenta, neglecting initial-state radiation.
The charged lepton is the selected isolated lepton, and the neutrino momentum is taken  as the missing momentum.
The $b$ quarks are identified with the two $b$-tagged jets among the four leading ones and the $W$ decay quarks with the two remaining ones.
The squared amplitudes are summed over helicities, over both assignments of the $b$ jets, and over the two permutations of the light-quark jets.
Generator-level distributions are obtained by evaluating the same discriminants on the parton-level four-momenta of the $b$ quarks and of the $W$ decay products, for the same selected events.

\begin{figure}[t]
\centering
\includegraphics[width=0.5\textwidth]{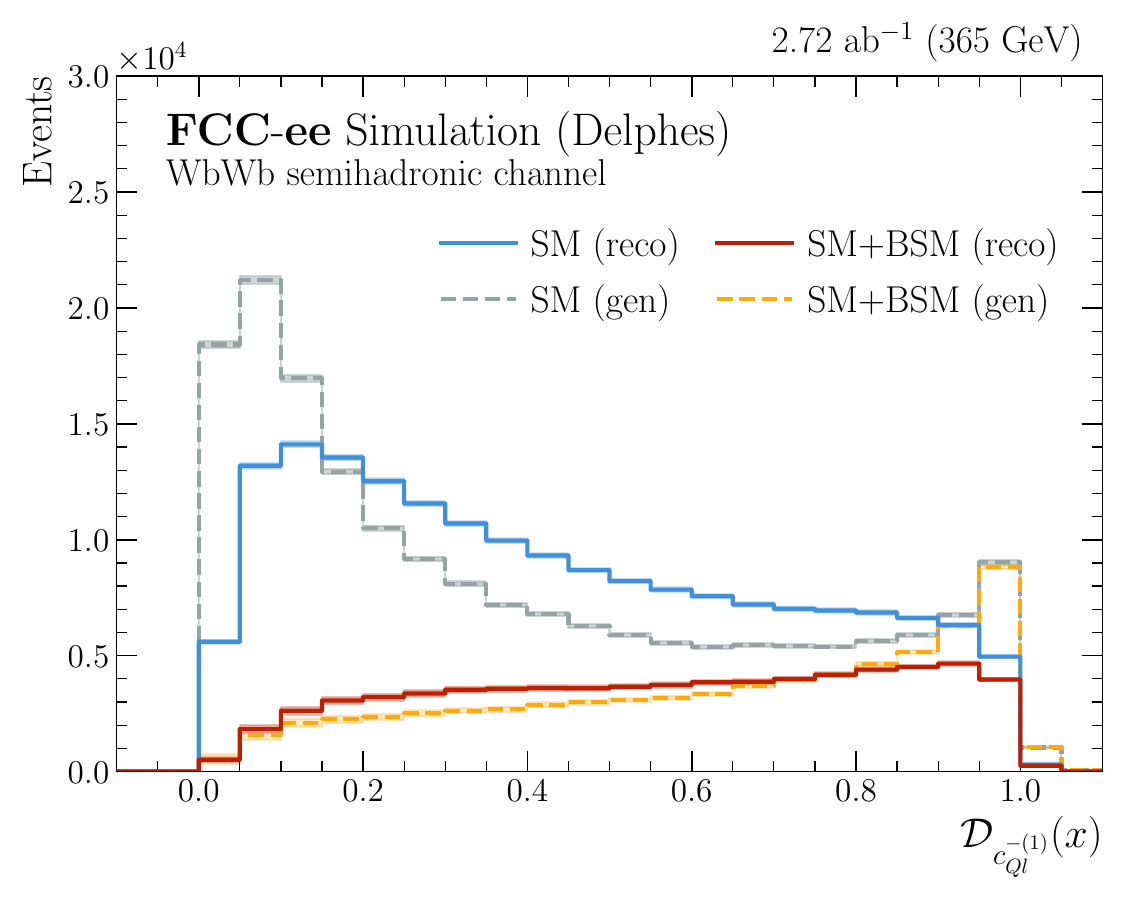}%
\includegraphics[width=0.5\textwidth]{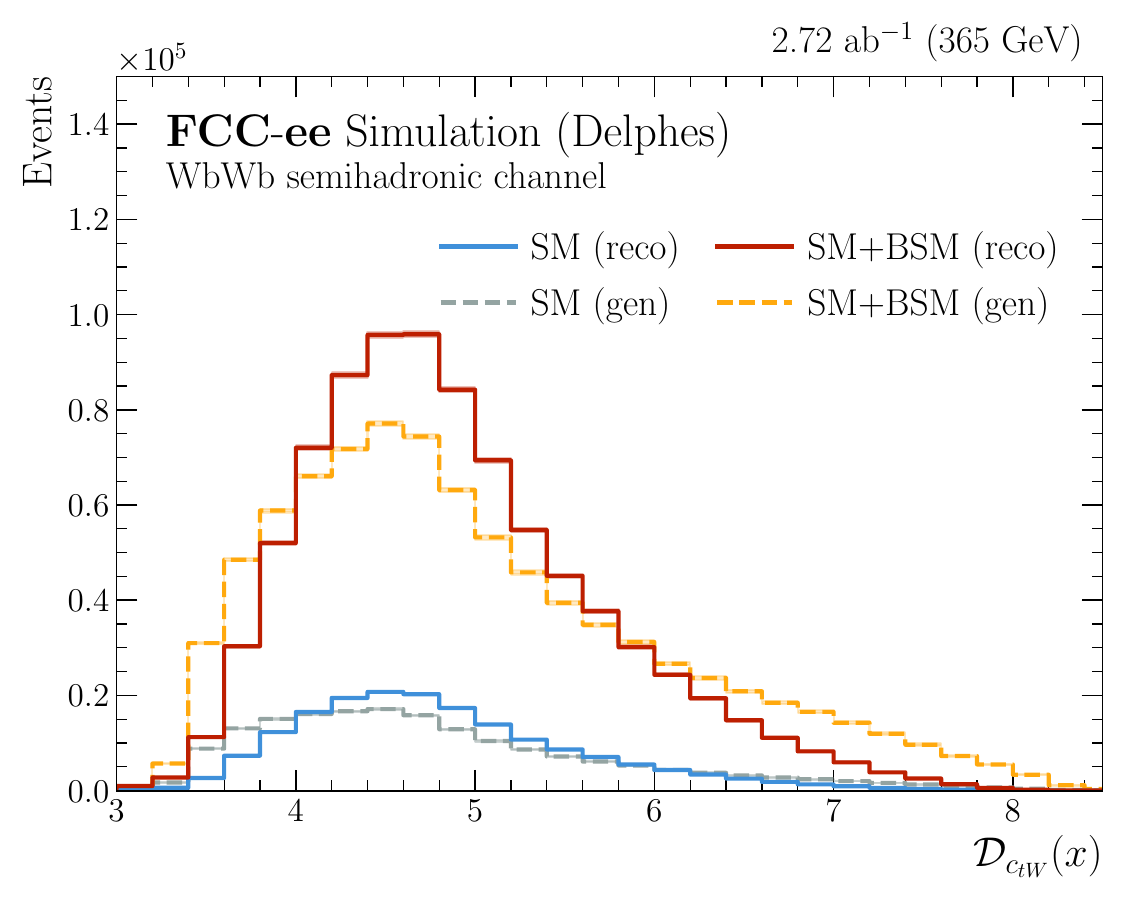}%
\caption{Distributions of the $\mathcal{D}_{\cQlM}$ (left) and $\mathcal{D}_{\ctW}$ (right) discriminants evaluated on SM and SMEFT samples with the corresponding $c_i/\Lambda^2$ coefficient set to $1/\text{TeV}^2$.
Dashed curves are obtained at the generator level and solid ones at the reconstruction level.
}
\label{fig:d_distribution}
\end{figure}

Neglecting detector effects and backgrounds, $\mathcal{D}$ is an optimal discriminant between the SM and the SMEFT hypothesis at the $\{c_i/\Lambda^2\}$ point.
Instead of scanning the multidimensional $\{c_i/\Lambda^2\}$ parameter space to achieve statistical optimality at every point as in the matrix-element method, we adopt a more economical procedure inspired by the optimal observable approach~\cite{Atwood:1991ka, Diehl:1993br}.
At linear order in the parameters of interest, the use of a single optimal observable per degree of freedom is known to achieve statistical optimality.
Retaining quadratic dependencies, we similarly define one discriminant per operator coefficient, obtained by setting a single $c_i/\Lambda^2$ to $1/\text{TeV}^2$ and all others to zero in Eq.~\eqref{eq:discriminant}.
While an optimal observable approach prescribes to only use the average value of each discriminant over the event sample, we consider their binned distribution, which is more robust against outliers.
Several such distributions are shown in Figure~\ref{fig:d_distribution}.
In the left panel, the coefficient $\cQlM$ is seen to visibly affect the shape of the corresponding $\mathcal{D}_{\cQlM}$ discriminant, while in the right panel $\ctW$ mostly affects the normalisation of the $\mathcal{D}_{\ctW}$ distribution.

The impact of reconstruction effects is visible in the difference between dashed distributions, obtained from generator-level kinematics, and solid ones, obtained after reconstruction, for the same selected events.
The dominant causes are the averaging over the four different permutations of assignment of $b$ jets and light-quark jets.
The estimate of the neutrino momentum from the missing momentum, biased by initial-state radiation photons lost along the beam axis, the jet energy response and resolution, including the energy carried away by neutrinos in semi-leptonic heavy-flavour decays, and the assignment of particles to jets and of jets to partons.
A further dilution arises from the other $W^+bW^-\bar{b}$ decay channels passing the selection, about 10\% of the sample, dominated by $W\to\tau\nu$ with a leptonic $\tau$ decay.
At the generator-level the $\tau$ lepton and its neutrino are used as input for the charged lepton and neutrino in the matrix element calculation.
At reconstruction level, the additional neutrinos from the $\tau$ decay degrade the kinematics, so that these events smear the distributions rather than bias them.
Single-top and non-resonant contributions, not described by the doubly resonant matrix elements, dilute the discrimination power at both levels.
Finally, the amplitudes are evaluated at the nominal rather than at the reduced effective centre-of-mass energy, since initial-state radiation is not reconstructed.
Appendix~\ref{app:components} presents the discriminant distributions, separating the $W^+bW^-\bar{b}$ signal contributions from those of other $W^+bW^-\bar{b}$ decays channels.

\begin{figure}[tb]
\centering
\includegraphics[width=0.6\linewidth]{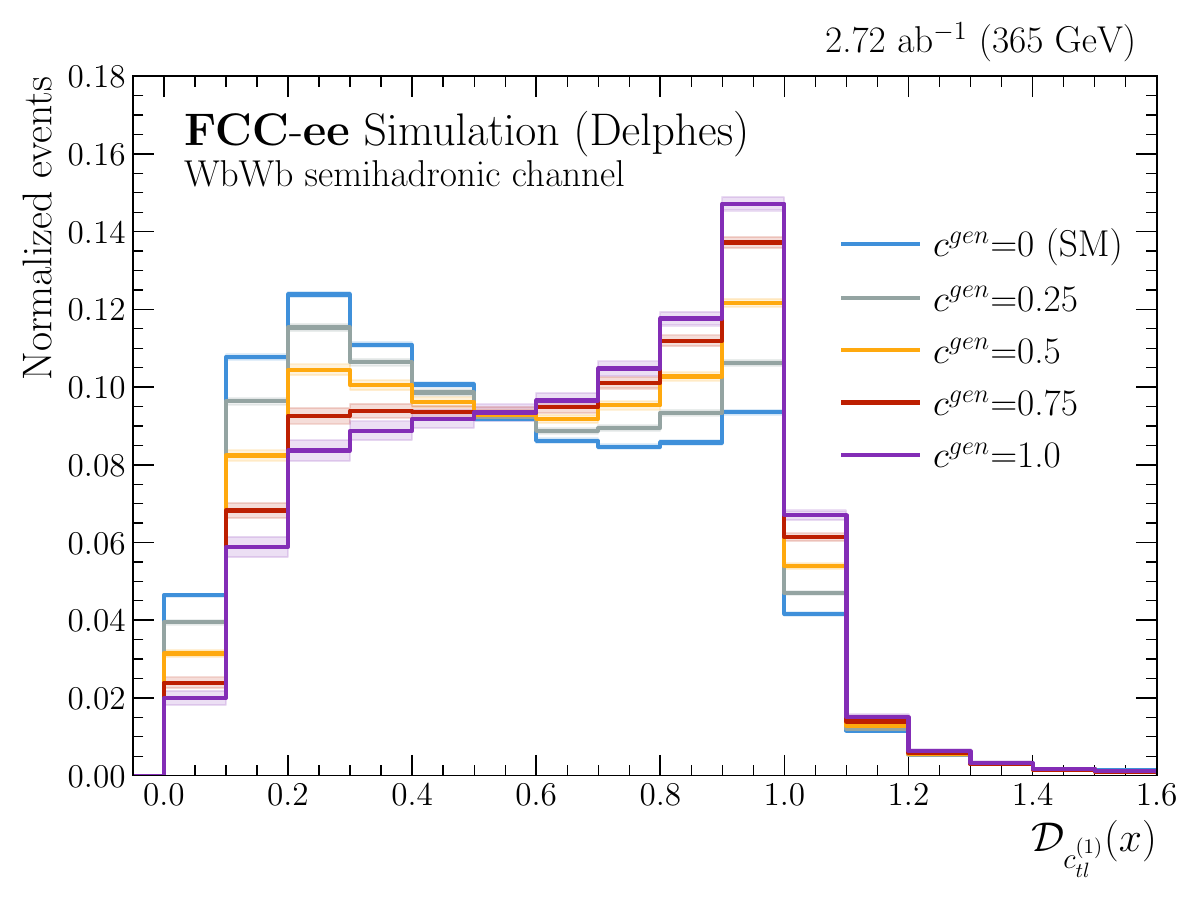}

\caption{Normalised shapes of the $\mathcal{D}_{\ctl}(x)$ discriminant evaluated on SMEFT samples generated with various values of $\ctl$.}

\label{fig:distributionvariation}
\end{figure}

To set global limits simultaneously on all $12$ operator coefficients considered, the $12$-dimensional space of multidimensional discriminants is binned.
To keep the total number of bins tractable, we adopt a minimal binning strategy in each dimension.
We note that most of the relevant information can be retained by just defining two bins per discriminant direction.
This can be understood by examining how the discriminant distributions vary when evaluated on SMEFT samples generated with different operator coefficient values.
Figure~\ref{fig:distributionvariation} for instance shows that variations of $\ctl$ make the $\mathcal{D}_{\ctl}$ discriminant pivot around a fixed crossing point.
The definition of just two bins, with a boundary at this specific point, therefore captures most of the sensitivity to the coefficient variation.
This behaviour is observed in most of the discriminants, and the bin boundaries are in practice set to the crossing point between the SM distribution and that obtained with a coefficient of $1/\text{TeV}^2$.

\section{Results}
\label{sec:results}

With the binned discriminants defined above, constraints on the various operator coefficients can be derived to assess the expected sensitivity, assuming the data follow the SM prediction.
A binned likelihood function is constructed as the product of Poisson probabilities over all bins, where the expectation value $\lambda(\{c_i\})$ in each bin is the sum of the SM contribution and the linear, quadratic, and cross-operator contributions for a given set of operator coefficients $\{c_i\}$.
The templates include the semi-leptonic $\WbWb$ signal and the other \WbWb\ final states passing the selection, both of which carry the dependence on the operator coefficients through the reweighting described in Section~\ref{sec:eventsim}.
Confidence intervals on the operator coefficients are extracted using a profile likelihood ratio test statistic $t = -2 \ln(\mathcal{L} / \mathcal{L}_{\max})$.

Systematic uncertainties are expected to be subdominant at the FCC-ee and as such are not included. The residual non-$\WbWb$ contamination amounts to about 1\% of the selected sample and its discriminant distributions differ substantially from the SMEFT-induced variations. 
Since the analysis is template based, with detector level distributions defining both the expected yields and the bin boundaries, migrations are consistently accounted for and residual experimental effects reduce to shape uncertainties on the templates, dominated by the jet energy scale and resolution and by the flavour tagging calibration. The large $W^+W^-$ and radiative return $Z$ samples collected at the top-pair energies, together with the lower energy FCC-ee runs, are expected to constrain these at the permil level, well below the statistical uncertainties. Finally, the fit assumes that the SM \WbWb\ prediction is known to better than 0.5\%. Since the same dataset serves for the extraction of these parameters and of the operator coefficients, a simultaneous determination would be the ultimate treatment and is left for future work. As a check, performing the global fit with a conservative 1\% uncertainty on the \WbWb\ normalisation impacts the global limits by less than 1\%.

\begin{figure}[tb]
\centering
\includegraphics[width=0.95\linewidth, trim=40 20 50 30, clip]{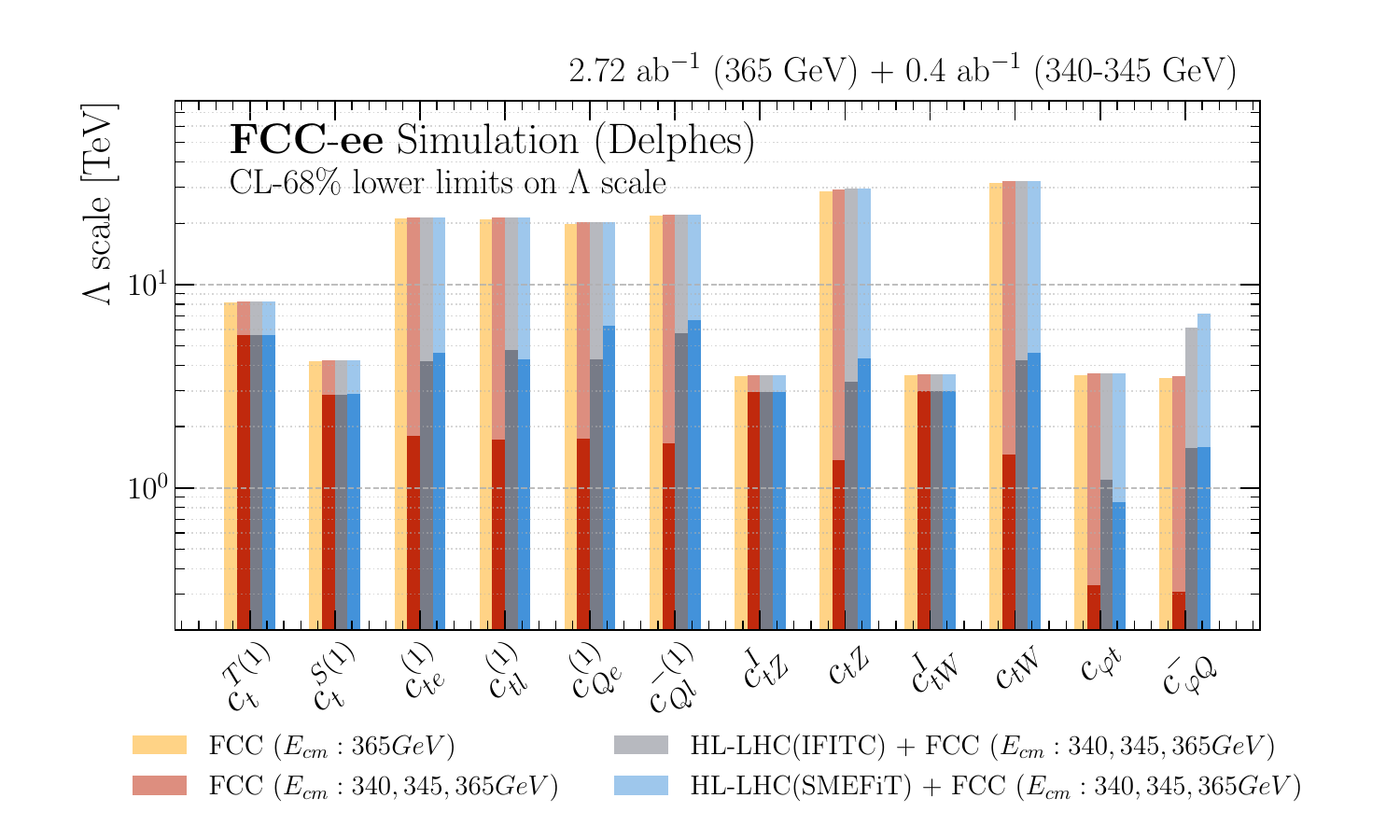}
\caption{Expected $68\%$ CL lower limits on the new physics scale $\Lambda/\sqrt{|c_i|}$.
The limits are shown with and without the HL-LHC constraints and the top-threshold datasets.
Global limits are shown as bars of darker shades, while individual limits are represented as bars of lighter shades.
}
\label{fig:limits}
\end{figure}

Figure~\ref{fig:limits} summarises our results.
Individual limits (lighter shades) are obtained by floating one coefficient at a time while fixing the others to zero, using approximately 60 uniform bins for the corresponding discriminant distribution.
Global limits (darker shades) are obtained by floating all coefficients simultaneously, using the 12-dimensional distribution of all discriminants with just two bins per dimension.

\begin{figure}[tb]
\centering
\includegraphics[width=0.6\linewidth]{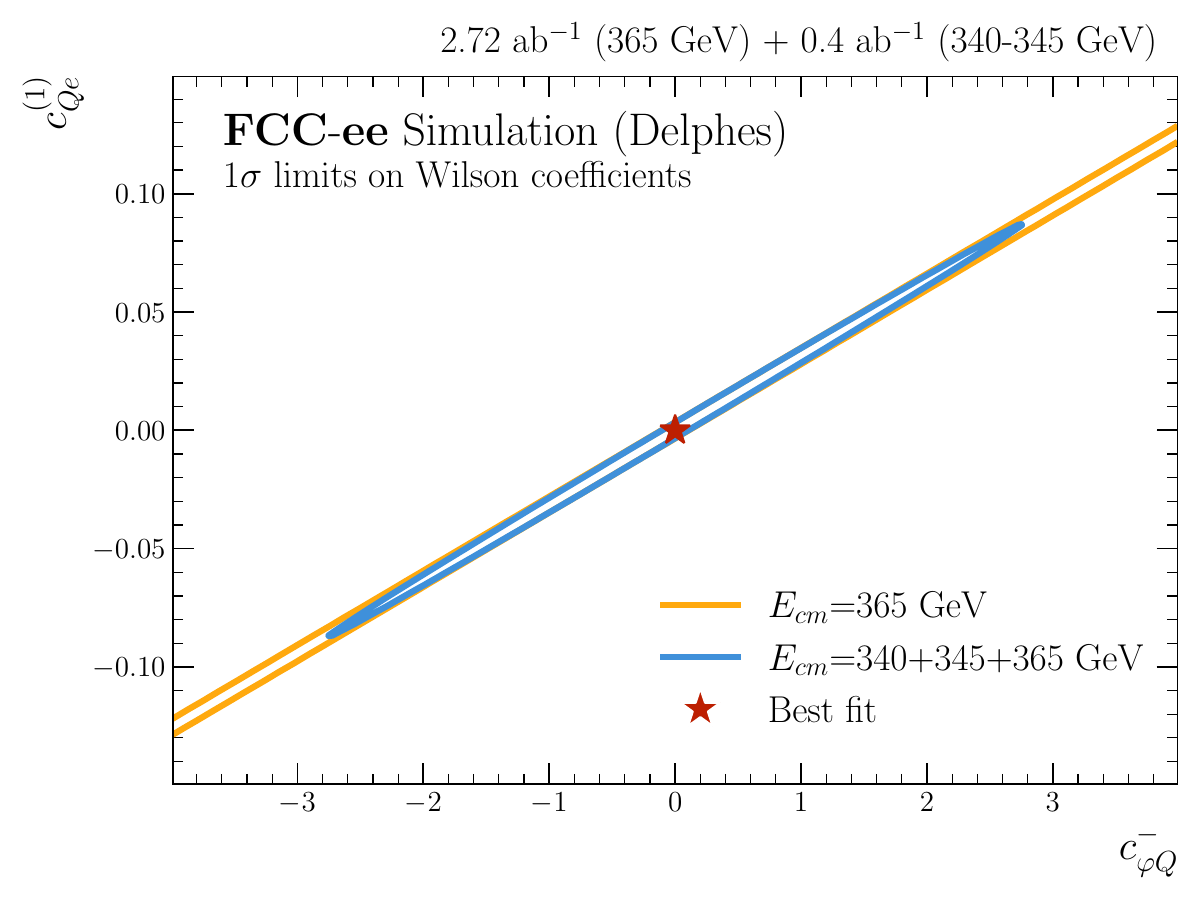}
\caption{Constraints in the ($\cQlM$, $\cQe$, $\PQqM$) space projected onto the ($\cQe, \PQqM$) plane.
The degeneracy remaining with $e^+e^-\to t\,\bar{t}$ measurements at a single centre-of-mass energy is lifted by the threshold scan data.}
\label{fig:scan2d}
\end{figure}

Most of the individual sensitivity stems from the $e^+e^-\to t\,\bar{t}$ measurement at \sqrtsTop.
That single energy run is however insufficient to simultaneously constrain all $12$ directions in a global fit: as discussed in Ref.~\cite{Durieux:2018tev}, the two-fermion and four-fermion operators with vector Lorentz structures are degenerate at a single centre-of-mass energy.
The top-threshold data at $\sqrt{s}=340-345\:$GeV allows to lift these degeneracies.
This is illustrated in Figure~\ref{fig:scan2d} showing the projection of a three-dimensional fit of $(\cQlM,\,\cQe,\,\PQqM)$ coefficients onto the $(\cQe,\,\PQqM)$ plane.
Beam polarisation would help disentangle the four-fermion operators involving left- and right-handed electrons, but would still leave a degeneracy that can only be lifted with distinct centre-of-mass energies.
The interplay between SMEFT constraints and the extraction of the top-quark mass and width planned in the threshold scan has not yet been studied in the literature.

A complementary handle to resolve these degeneracies could be to exploit the initial-state radiation of a photon to effectively access lower $e^+e^-\to t\,\bar t$ centre-of-mass energies.
The emitted photon would need to be reconstructed to determine the effective event-by-event $e^+e^-\to t\,\bar t$ centre-of-mass energy.
Statistical uncertainties could however be limiting.
The assessment of the potential of such an approach is left to future work.

\begin{figure}[tb]
\adjustbox{width=\textwidth}{
\begin{tabular}{@{}c@{\;}c@{}}
\raisebox{0mm}{\includegraphics{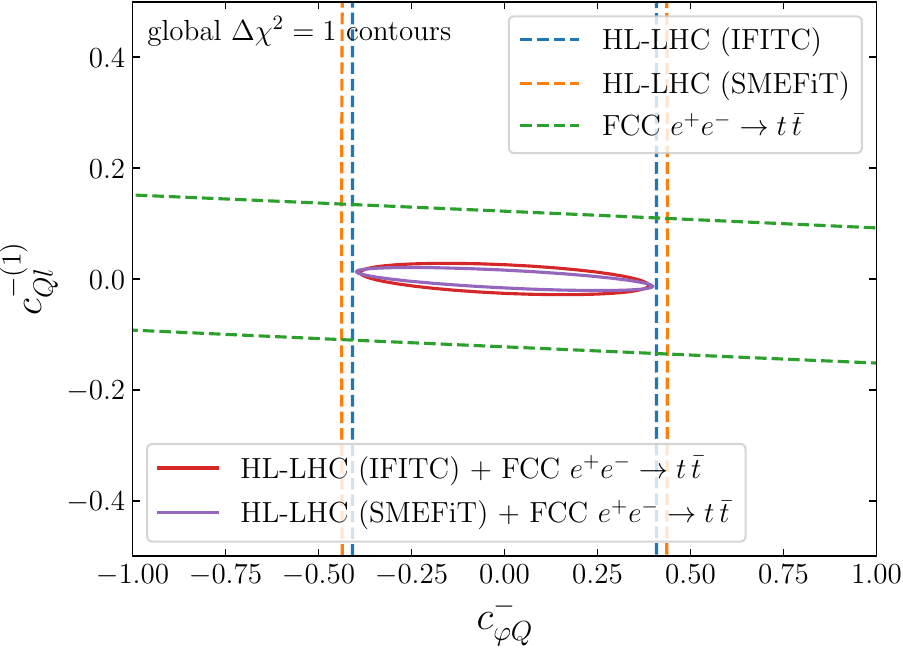}}
&\includegraphics{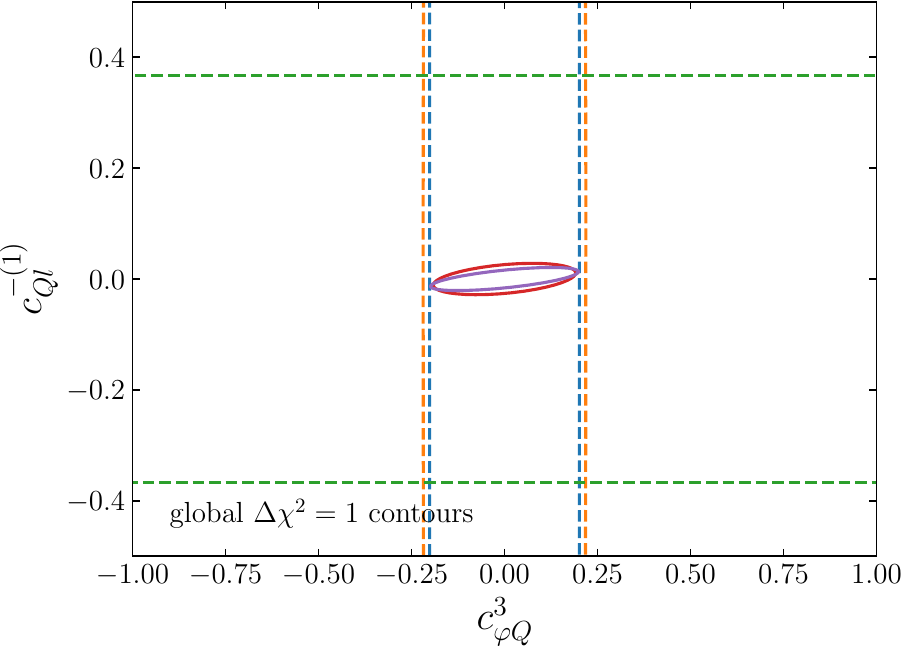}
\\
\raisebox{-3.5mm}{\includegraphics{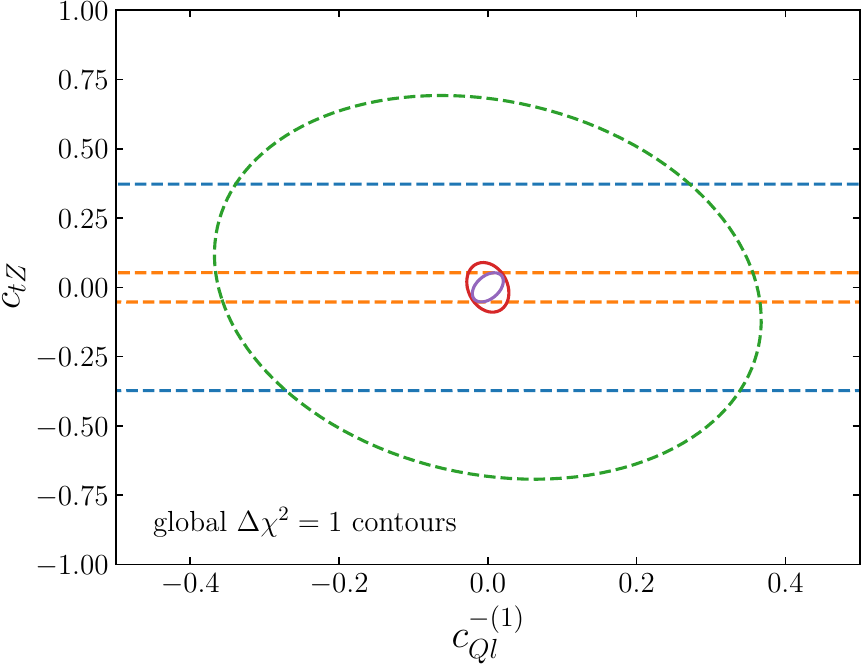}}
&\includegraphics{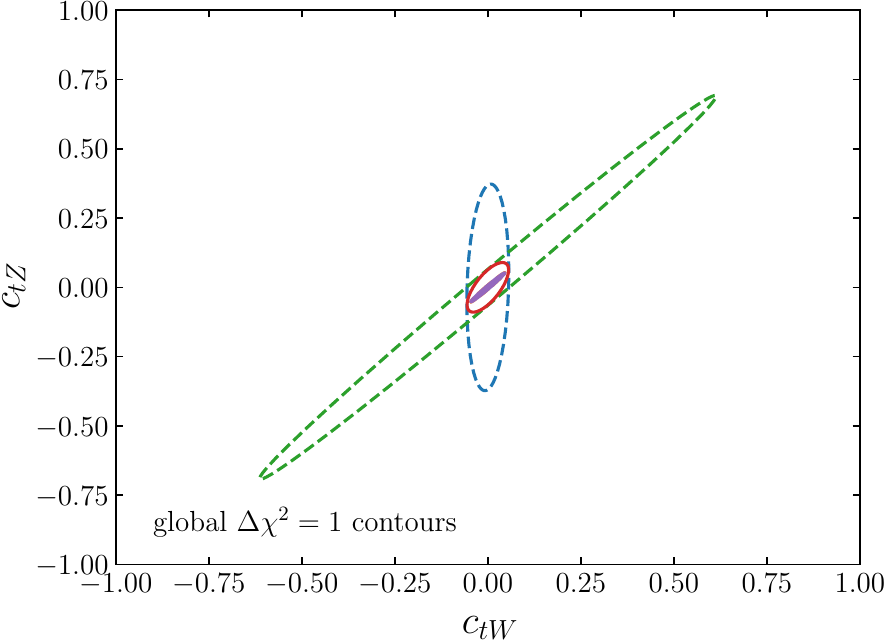}
\end{tabular}%
}
\caption{Two-dimensional projections of global linear sensitivities, in terms of $\Delta\chi^2=1$ contours, to top-quark operators.
The HL-LHC and FCC $e^+e^-\to t\,\bar{t}$ prospects are shown in isolation, and combined.
}
\label{fig:2d-projections}
\end{figure}

For a more complete assessment of the top-quark operator constraints after the $e^+e^-\to t\,\bar t$ measurements at FCC-ee, we also incorporate in our fit the prospective HL-LHC constraints compiled by the IFITC~\cite{Cornet-Gomez:2025jot} and SMEFiT~\cite{Armadillo:2026mvp} collaborations.
We use the covariance matrix obtained in a linear fit approximation and do not include renormalisation-group running to a high scale, which would blur the separation between the top-quark sector and the others.\footnote{The limited numerical accuracy provided on the HL-LHC correlation matrix in Fig.~12 of Ref.~\cite{Cornet-Gomez:2025jot} leads to negative eigenvalues which we regulate by taking their absolute value.
We verify the resulting regularised covariance matrix reproduces the global constraints obtained in Ref.~\cite{Cornet-Gomez:2025jot}.}
All the operator coefficients not considered in our analysis are marginalised over.
The CP-odd electroweak dipole operator coefficients, $\ctWI$ and $\ctZI$, as well as the scalar and tensor four-fermion operator coefficients, $\ctlS$ and $\ctlT$, are not included in IFITC and SMEFiT studies \cite{Cornet-Gomez:2025jot,Armadillo:2026mvp}.
The $\ctl$ and $\cte$ four-fermion operator coefficients are loosely constrained, mostly through $pp\to t\,\bar{t}\,e^+e^-$ measurements that are not necessarily optimised for that purpose.
The extrapolations of dedicated measurements to the HL-LHC sensitivity~\cite{Collaboration:2938605} however still predict a sensitivity about one order of magnitude weaker than our results.	
The $\cQlM+2\cQla$ and $\cQe$ combinations of four-fermion operator coefficients are better probed in $b\,\bar{b}\to e^+e^-$ scattering at the LHC and in $e^+e^-\to b\,\bar{b}$ at LEP, than in top-quark processes.
These combinations of four-fermion operator coefficients would also be tightly constrained by FCC-ee measurements of $b\,\bar{b}$ pair production at high centre-of-mass energies, which are however not included in our top-centric analysis.
The IFITC and SMEFiT bounds on our parameter space have comparable strengths and only mild differences remain after combination with our $e^+e^-\to t\,\bar t$ measurements (see Figure~\ref{fig:2d-projections}).

This combination of $e^+e^-\to t\,\bar t$ prospects with HL-LHC ones only improves the individual constraints for the $\PQqM$ coefficient.
The $\PQqM+2\PQqa$ linear combination is indeed precisely probed by the LEP measurement of $Z\to b\bar{b}$, which is included in the HL-LHC projections of IFITC and SMEFiT.
The $Z$-pole run of FCC-ee which would further tighten this constraint is not included in the present analysis since we focus just on $e^+e^-\to t\,\bar t$.

The combination with HL-LHC prospects improves the global constraints much more broadly.
It brings complementary sensitivity and, in particular, helps to further reduce the approximate degeneracies between two- and four-fermion operators.
The two-dimensional projections shown in Figure~\ref{fig:2d-projections} highlight some of the complementarities between HL-LHC and FCC $e^+e^-\to t\,\bar{t}$ sensitivities.
For simplicity, these are obtained using a Gaussian covariance-matrix approximation of our FCC-ee prospects, which reproduces well both its individual and global constraints, but exclude scalar and tensor four-fermion operators whose dependencies in differential rate are only quadratic.
We provide this covariance matrix as ancillary file to facilitate the incorporation of our constraints into more global analyses and display the associated correlation matrix in Appendix~\ref{app:correlation}.

\begin{figure}[tb]\centering
\includegraphics[width=.5\textwidth]{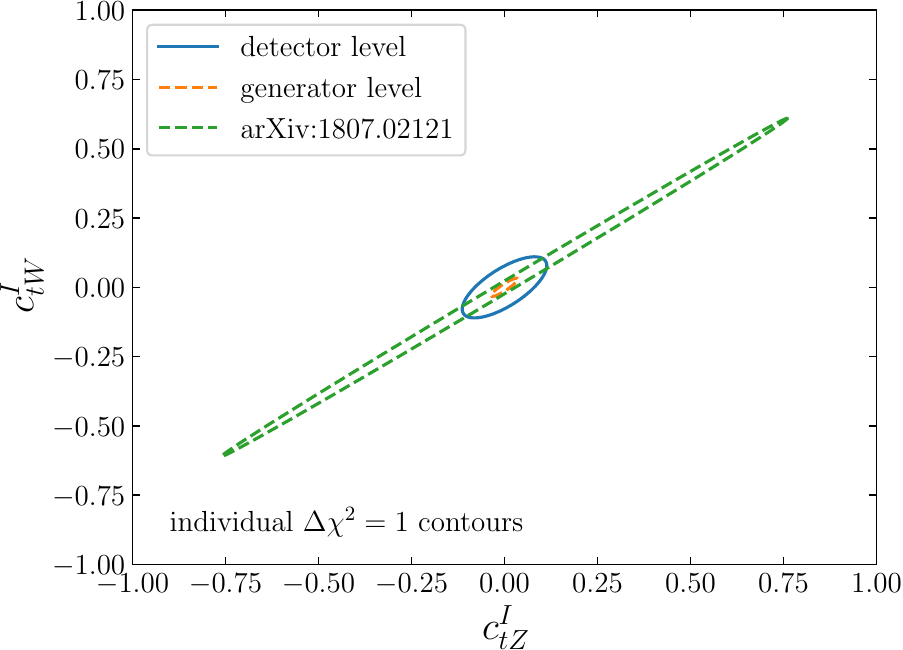}%
\includegraphics[width=.5\textwidth]{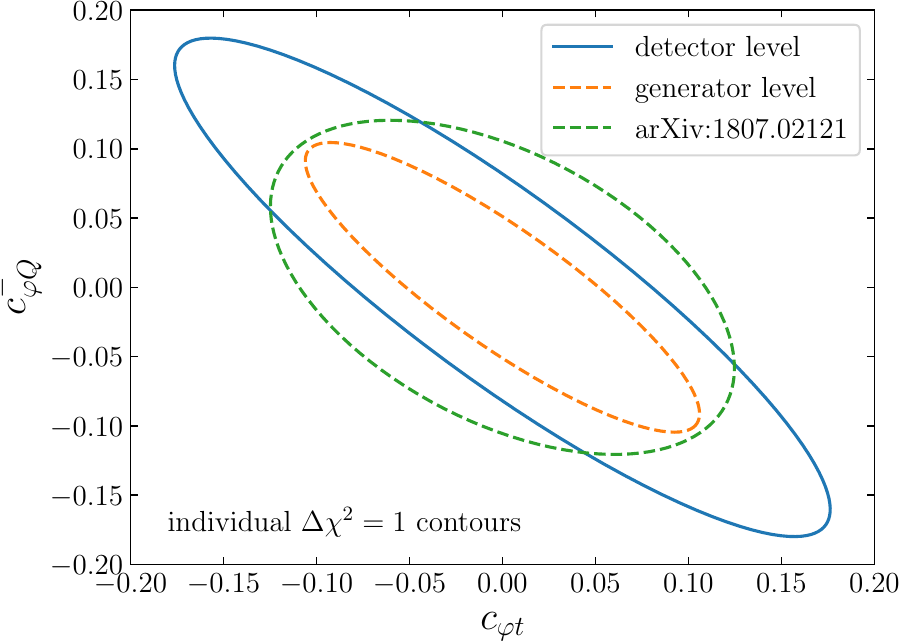}%
\caption{Individual two-dimensional slices of the SMEFT parameter space comparing the results of our analysis at the detector level (blue) and generator level (orange) with those of Ref.~\cite{Durieux:2018tev} (green), in terms of $\Delta\chi^2=1$ contours.
}
\label{fig:reco-gen-old-individual}

\includegraphics[width=.5\textwidth]{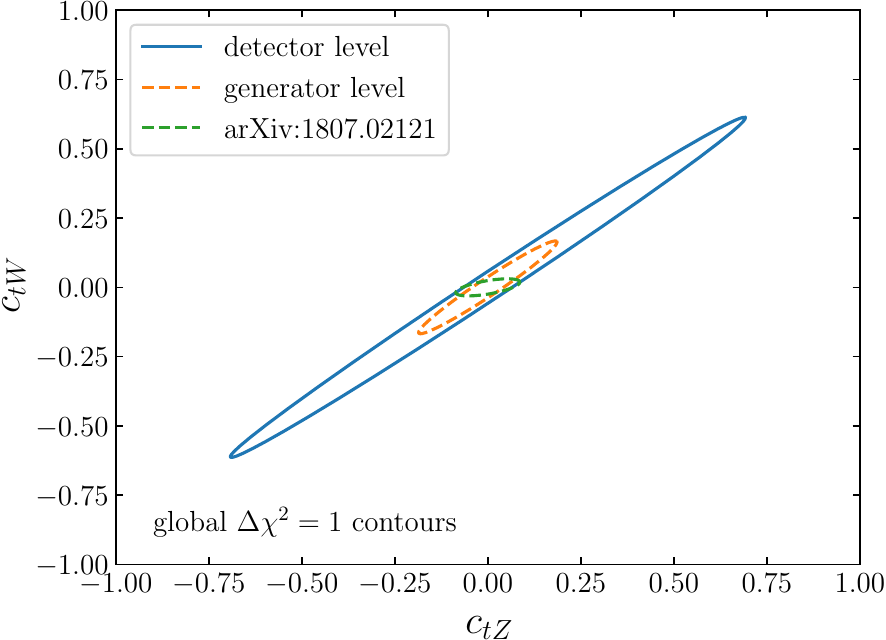}%
\includegraphics[width=.5\textwidth]{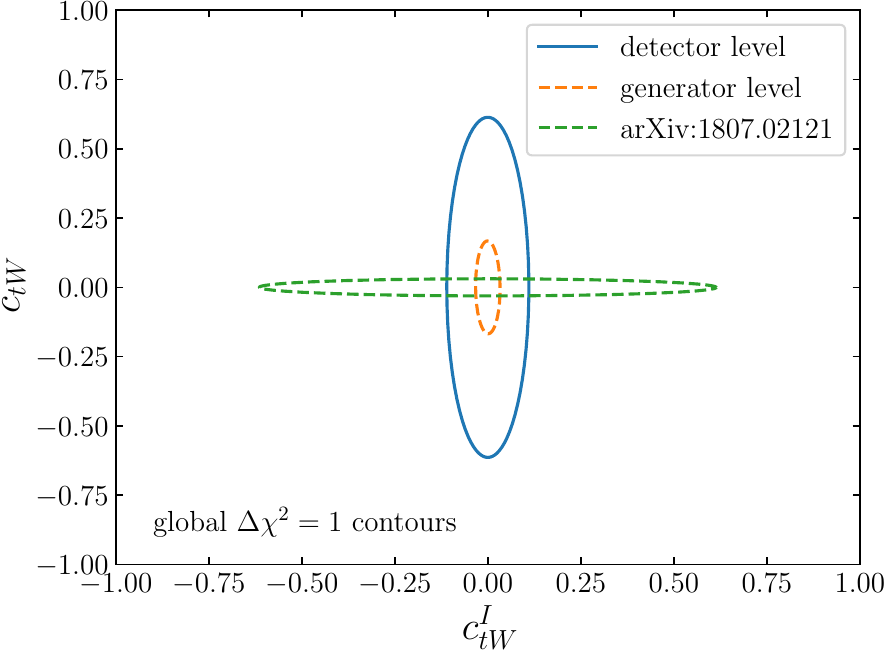}%
\caption{Global two-dimensional projections in the SMEFT parameter space comparing the results of our analysis at the detector level (blue) and generator level (orange) with those of Ref.~\cite{Durieux:2018tev} (green), in terms of $\Delta\chi^2=1$ contours.
}
\label{fig:reco-gen-old-global}
\end{figure}

In comparison with that of Ref.~\cite{Durieux:2018tev},\footnote{Using the results of Ref.~\cite{Durieux:2018tev} available at \url{https://github.com/gdurieux/optimal_observables_ee2tt2bwbw}, the luminosities have been increased to $0.4\:\text{ab}^{-1}$ at $\sqrt{s}=350\:$GeV and $2.72\:\text{ab}^{-1}$ at $\sqrt{s}=365\:$GeV to approximately match the run scenario considered here.} the present analysis gains constraining power thanks to the additional kinematic information of the $W$ decay products included in the discriminant definition.
In individual two-dimensional slices of the parameter space (see examples in Figure~\ref{fig:reco-gen-old-individual}), this is especially effective in probing the least constrained linear combination of $\ctWI$--$\ctZI$ pair of CP-odd electroweak dipole operator coefficients.
It also helps reducing some of the strongest correlations appearing in these two-dimensional slices, which involve the CP-even $\ctW$ and $\ctZ$ dipole operator coefficients.
On the contrary, our more realistic treatment of detector effects smears distributions and diminishes the discriminating power between different operators.
This notably makes it more difficult to distinguish operators that only differ by the chirality of the top quarks they involve, increasing correlations between the $\ctl$--$\cQlM$, $\cte$--$\cQe$, and $\PQqM$--$\cpt$ pairs.
At the global level (see examples in Figure~\ref{fig:reco-gen-old-global}), the higher level of approximate degeneracies due to detector effects especially affects the $\ctW$ and $\ctZ$ constraints which become looser than those obtained in Ref.~\cite{Durieux:2018tev}.
Since the $\ctWI$--$\ctZI$ pair has essentially no correlation with other operator coefficients, the extra individual constraining power brought by the additional kinematic information is mostly preserved at the global level, leading to constraints notably better than in Ref.~\cite{Durieux:2018tev}.
In directions other than these four electroweak dipole operator coefficients, the moderate gain of individual constraining power and moderate loss of discriminating power balance each other to lead to global constraints comparable to those of Ref.~\cite{Durieux:2018tev}.
The global combination with HL-LHC prospects preserves these qualitative features.


\section{Conclusion} \label{sec:Conclusion}
In this paper, we explored the sensitivity of top-quark pair production at the FCC-ee to dimension-six SMEFT operators.
Using a fast simulation of the IDEA detector concept, we showed that the clean leptonic $e^+ e^-$ environment provides an exceptionally pure $\WbWb$ sample, with a total background contamination at the percent level.
We constructed nearly optimal discriminants based on ratios of matrix elements, exploiting the full multi-dimensional kinematics of the process.
We confirmed that the threshold-scan data, despite its small luminosity, is important to lift the degeneracies between two-fermion and four-fermion operators that arise when considering a single centre-of-mass energy.
The realistic event selection needed to reduce backgrounds, the impact of event reconstruction on discriminant distributions, and the inclusion of the full six-body kinematic information in the discriminant construction are the most notable improvements compared to previous studies~\cite{Janot:2015yza, Durieux:2018tev}.
The binning scheme of multi-dimensional discriminant distributions is also novel, providing a simple and robust methodology, while maintaining a nearly optimal sensitivity.

While the fast simulation with \DELPHES provides a reliable representation of the IDEA detector response, it inherently simplifies certain experimental complexities.
The fast simulation notably relies on an idealised particle-matter description that neglects secondary emissions and instrumental particle misidentification.
Consequently, the presence of these instrumental effects could lead to a higher background contamination, that could be suppressed with tighter identification criteria at the cost of small losses in signal efficiency, and to degradation of the resolution of the reconstructed four-momenta of the objects entering the matrix-element discriminants, which are highly sensitive to the multi-body kinematics.
However, given the exceptionally clean $\mathcal{O}(1\%)$ background level achieved after the baseline selection and the minimal binning in the discriminants, these unmodelled effects are expected to have a subdominant impact on the ultimate SMEFT constraints. 

A possible future improvement to the current work is to incorporate transfer functions in the matrix-element-based discriminants to treat detector smearing consistently, include ISR reconstruction to assess the effective centre-of-mass energy of each event, and introduce a dedicated treatment of the $\tau$ channel, which currently
dilutes the discriminant shapes. Accounting for the SMEFT sensitivity of
single-resonant and non-resonant contributions to the $W^+bW^-\bar{b}$
final state could improve our constraints slightly.
The possible interplay between the top-quark mass, width, and electroweak coupling extraction in the threshold scan also remains to be investigated.

\acknowledgments

We thank our colleague Louis Portales for generating several of the Monte Carlo samples used in this analysis, and Matteo Defranchis and Ankita Mehta for providing the
code for the \ttbar event selection and analysis.
We are also grateful to Víctor Miralles and Luca Mantani for providing us with the HL-LHC fit covariances of IFITC and SMEFiT collaborations, respectively.
Finally, we thank Marcel Vos, Juergen Reuter, Xunwu Zuo, Tommaso Dorigo and Patrizia Azzi for their useful discussions, suggestions, and valuable feedback.
G.~D.\ is a Research Associate of the Fonds de la Recherche Scientifique -- FNRS, Belgium, supported in part by the IISN convention 4.4509.26.
S.~S.~C.'s work was supported by the ``Ram\'on y Cajal'' program under Project No.\ RYC2024-048719-I, funded by ICIU/AEI/10.13039/501100011033 and by the FSE+.

\appendix


\section{Reconstruction and background effects on the discriminants}\label{app:components}

\begin{figure}[tb]
\centering
\includegraphics[width=0.5\textwidth]{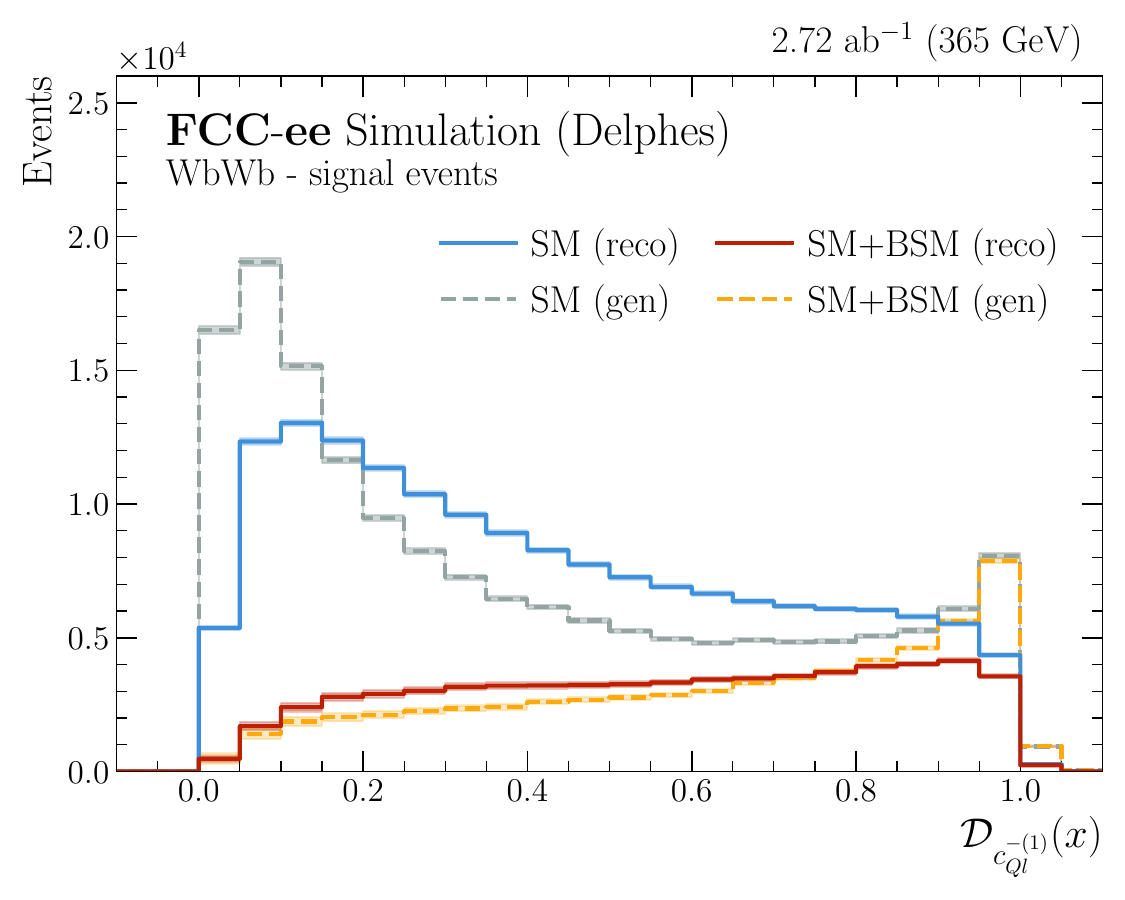}%
\includegraphics[width=0.5\textwidth]{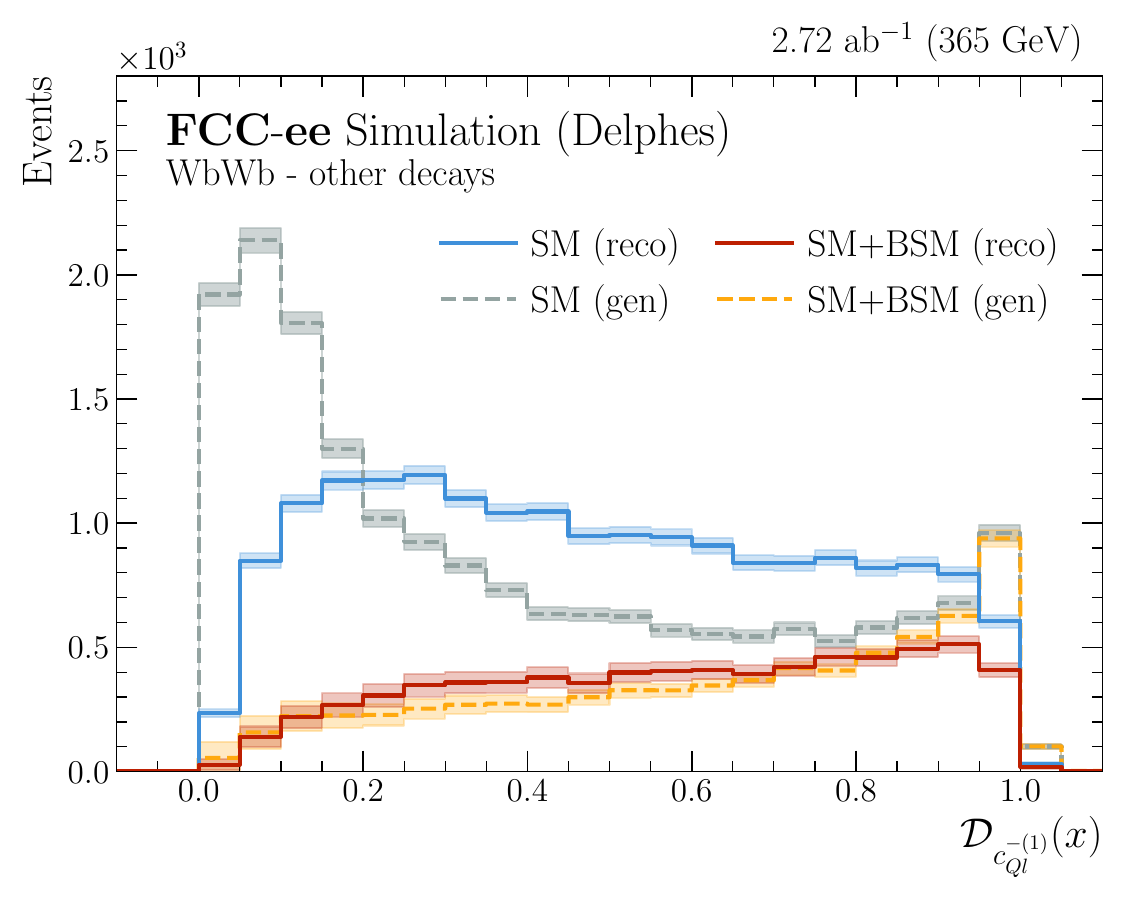}%
\vspace{0.2 cm}

\centering
\includegraphics[width=0.5\textwidth]{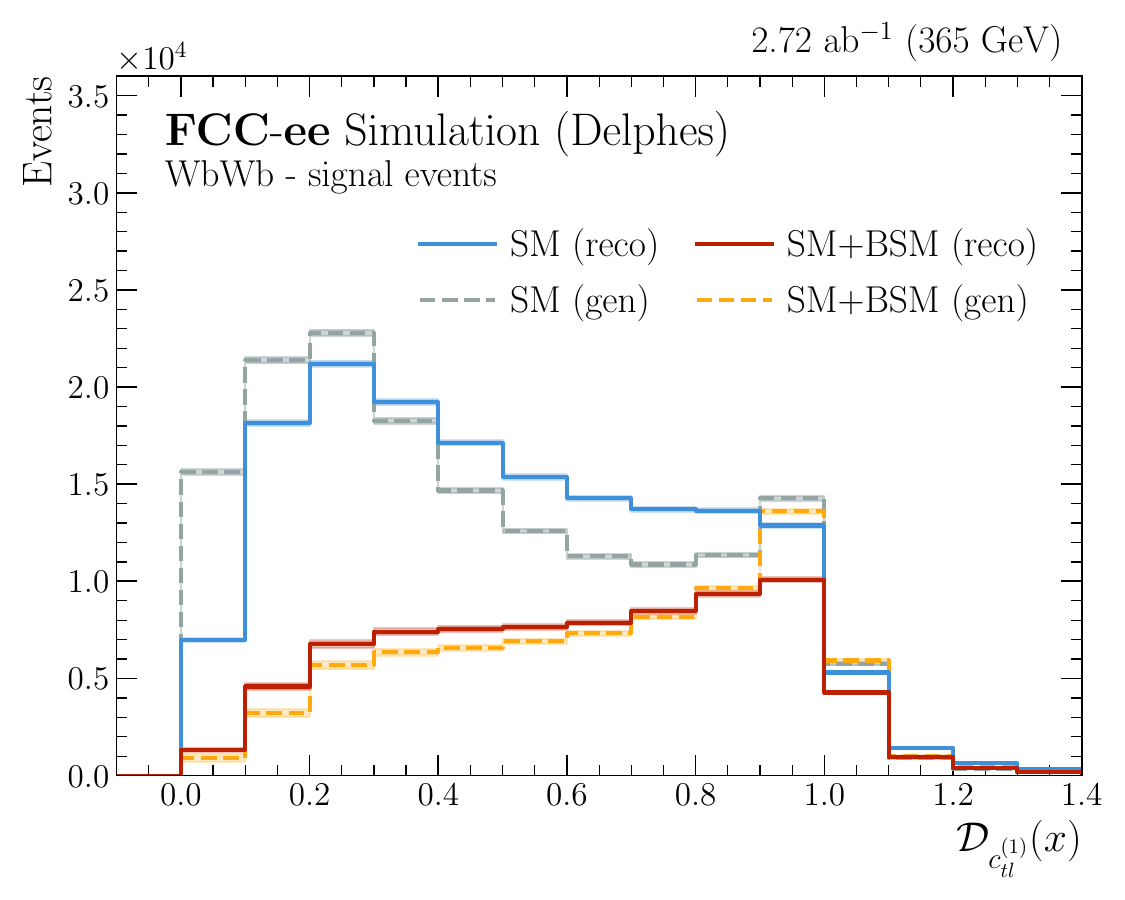}%
\includegraphics[width=0.5\textwidth]{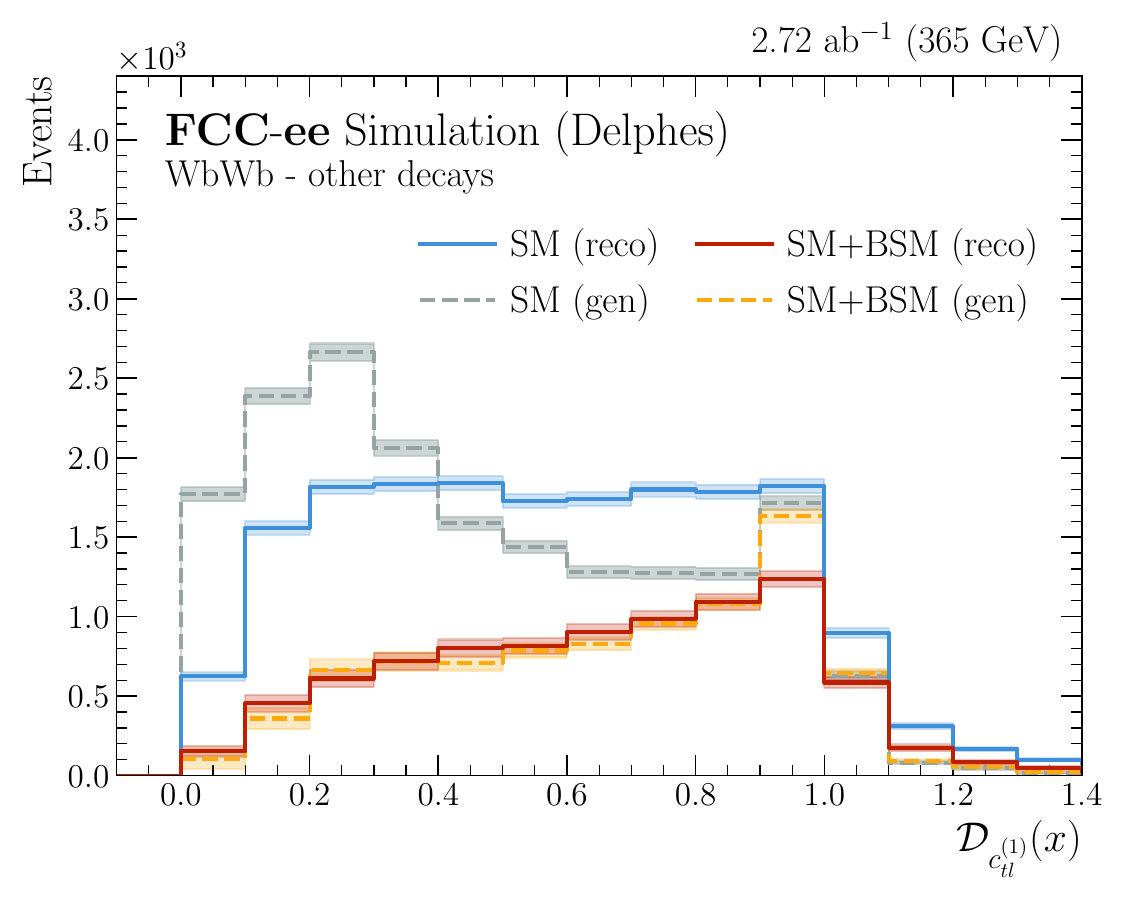}%

\vspace{0.2 cm}

\centering
\includegraphics[width=0.5\textwidth]{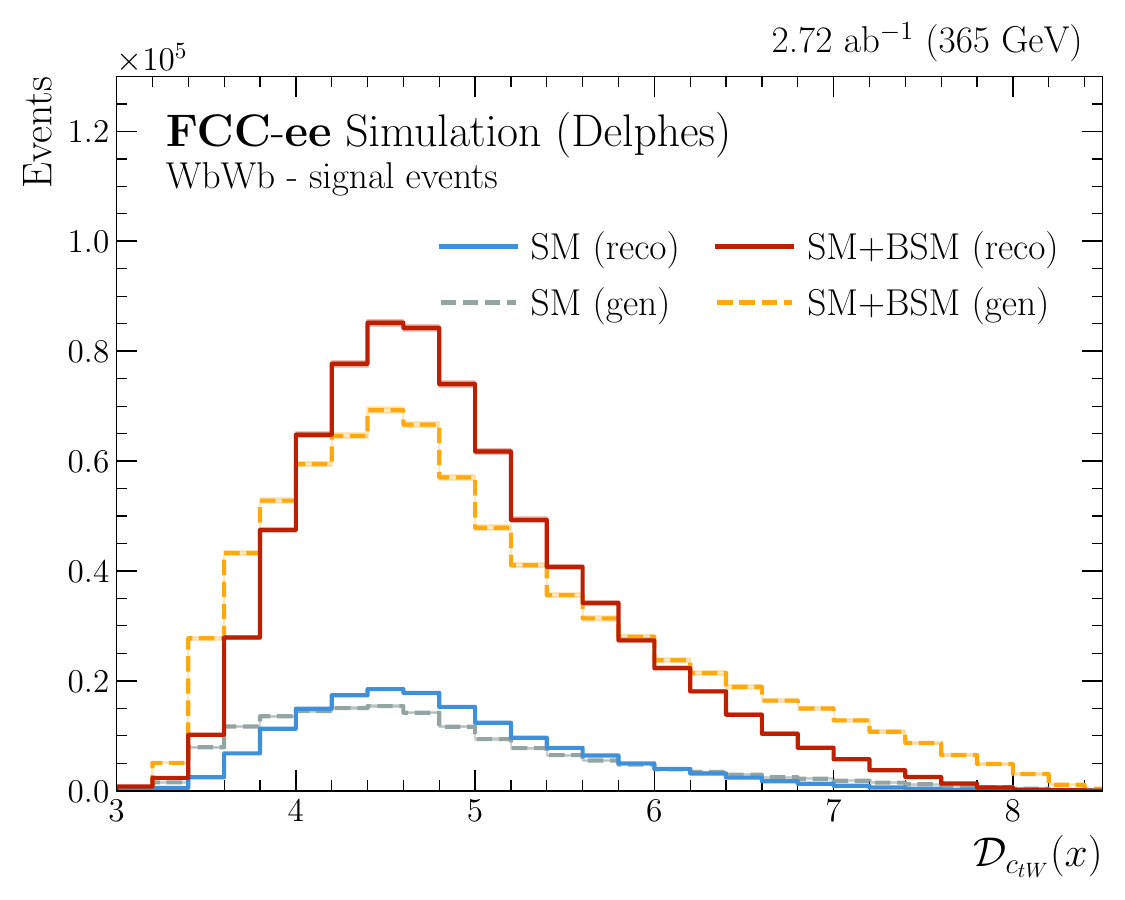}%
\includegraphics[width=0.5\textwidth]{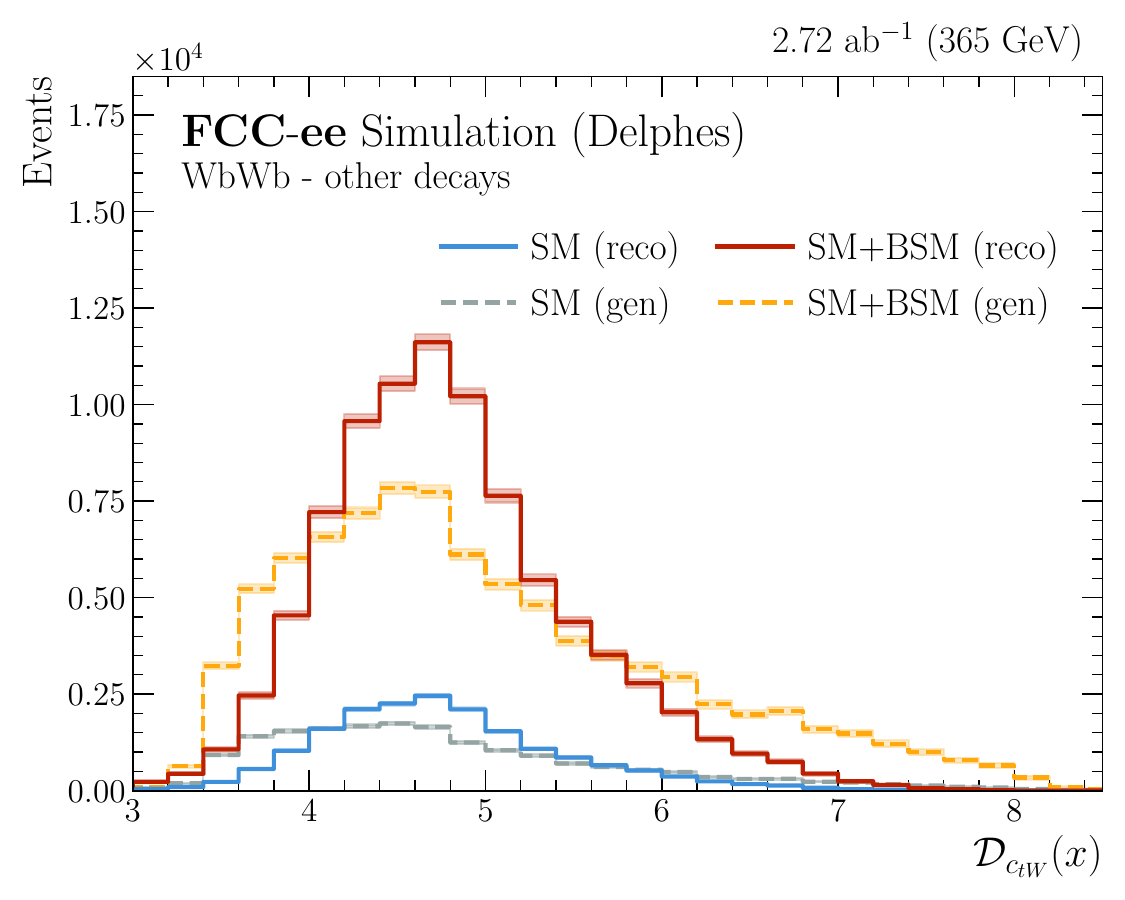}%
\caption{Distributions of the $\mathcal{D}_{\cQlM}$, $\mathcal{D}_{\ctl}$ and $\mathcal{D}_{\ctW}$ discriminants evaluated on SM and SMEFT samples with the corresponding $c_i/\Lambda^2$ coefficient set to $1/\text{TeV}^2$.
Dashed curves are obtained at the generator level and solid ones at the reconstruction level.
The $\WbWb$ signal and background events are separated in the left and right plots, respectively.
}
\label{fig:components}
\end{figure}

The degradation of the discriminant distributions at the reconstruction level originates from two main sources: pure detector reconstruction effects and contamination from alternative decay modes  different from the semi-leptonic $\WbWb$ in the $e$ or $\mu$ channel. To better separate these effects, in Figure~\ref{fig:components} the distributions for some discriminants are broken down into their constituent parts, isolating the $W^+bW^-\bar{b}$ signal events from the remaining $W^+bW^-\bar{b}$ background decays. For observables where a variation in the EFT coefficient primarily alters the shape of the discriminant, such as $\mathcal{D}_{\cQlM}$ and $\mathcal{D}_{\ctl}$, the degradation of discrimination power is significantly worse in the background events than in the signal ones. Conversely, for discriminants such as $\mathcal{D}_{\ctW} $,  where the EFT effect predominantly scales the overall event normalisation, this degradation is far less impactful.

\section{Correlation matrix}
\label{app:correlation}

\begin{figure}[!htbp]
    \centering 
    \includegraphics[width=0.9\textwidth]{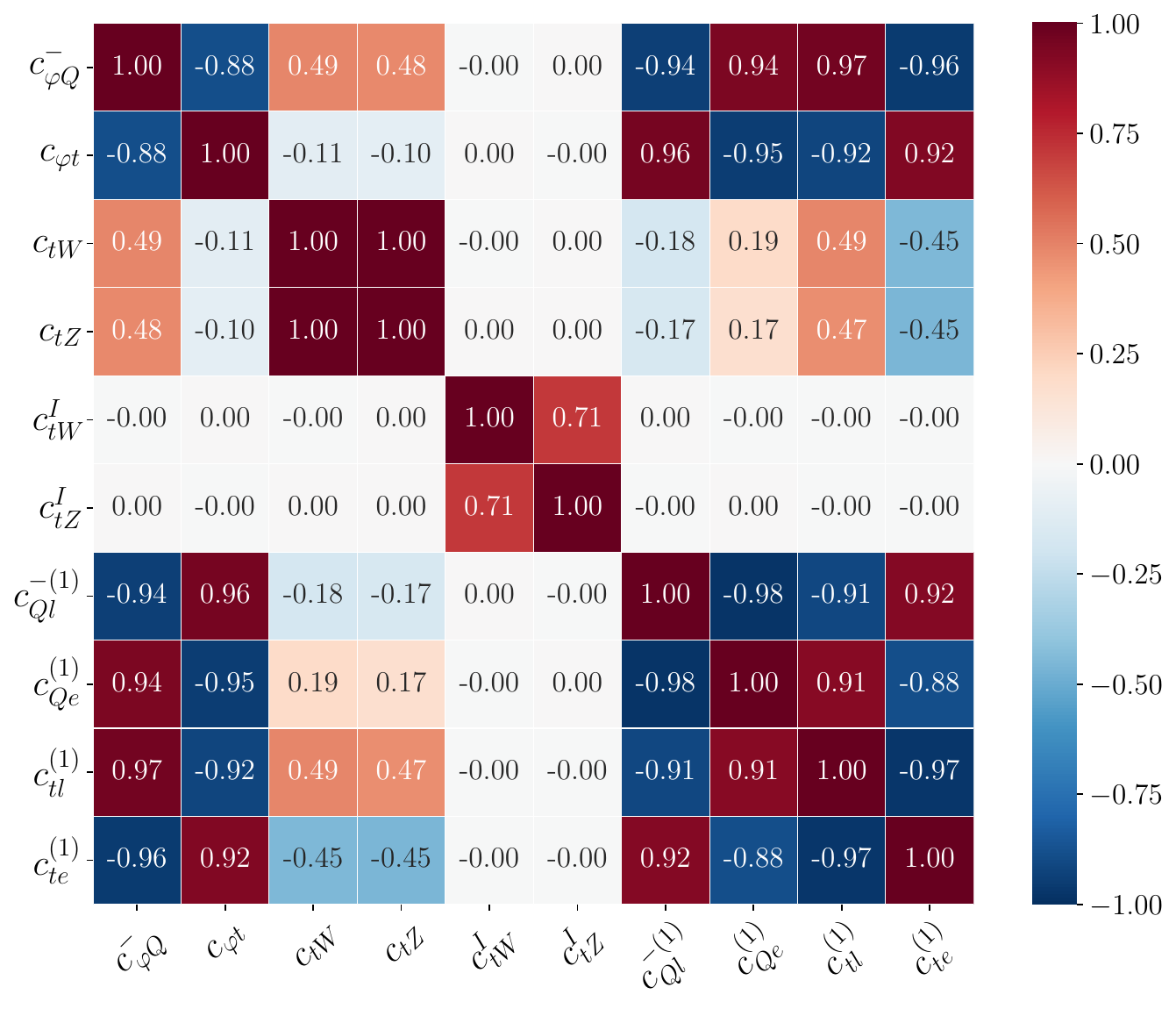}
  
    \caption{Correlation matrix of our $e^+e^-\to t\,\bar{t}$ fit at the FCC-ee.}
    \label{fig:correlation_matrix}
\end{figure}

The correlation matrix of our $e^+e^-\to t\,\bar{t}$ fit at the FCC-ee, including both \sqrtsTop and $\sqrt{s}=340$--345~GeV datasets, is shown in Figure~\ref{fig:correlation_matrix}.
The matrix is obtained from the Hessian of the $\Delta\log\mathcal{L}$ function evaluated at the SM point, corresponding to vanishing operator coefficients.
The Hessian is subsequently inverted and normalised to yield the correlation matrix.

The operators $c_t^{S(1)}$ and $c_t^{T(1)}$ are excluded from the estimation because they contribute only at quadratic order.
As a result, the likelihood is effectively flat in their directions in the vicinity of the minimum, preventing a reliable determination of their correlations with the remaining parameters.

\bibliographystyle{JHEP}
\bibliography{FCC}

\end{document}